# Hardness Amplification in Proof Complexity

Paul Beame[*]
Computer Science & Engineering
University of Washington
Seattle, WA 98195-2350
beame@cs.washington.edu

Trinh Huynh[*][†]
Computer Science & Engineering
University of Washington
Seattle, WA 98195-2350
trinh@cs.washington.edu

Toniann Pitassi[‡]
Computer Science
University of Toronto
Toronto, ON M5S 1A4
toni@cs.toronto.edu

October 31, 2018

**Abstract**

We present a general method for converting any family of unsatisfiable CNF formulas that is hard for one of the simplest proof systems, tree resolution (ordinary backtracking search), into formulas that require large rank in very strong proof systems, which include any proof system that manipulates polynomials or polynomial threshold functions of degree at most $k$ (known as Th($k$) proofs). These include high degree versions of Lovász-Schrijver systems, (LS($k$), LS$_+(k)$), high degree versions of Cutting Planes proofs (CP($k$)), Sherali-Adams, and Lasserre proofs.

We introduce two very general families of these very strong proof systems, denoted by $T^{cc}(k)$ and $R^{cc}(k)$. The proof lines of $T^{cc}(k)$ are arbitrary Boolean functions, each of which can be evaluated by an efficient $k$-party randomized communication protocol. $T^{cc}(k)$ proofs include Th($k-1$) proofs as a special case. $R^{cc}(k)$ proofs generalize $T^{cc}(k)$ proofs and require only that each inference be checkable (in a certain weak sense) by an efficient $k$-party randomized communication protocol.

Our main results are the following:

- For all $k \in O(\log \log n)$, any unsatisfiable CNF formula $F$ requiring resolution rank $r$ can be converted to a related CNF formula $G = \text{Lift}_k(F)$ requiring refutation rank $r^{\Omega(1/k)}/\log^{O(1)} n$ in all $R^{cc}(k)$ systems. Since resolution rank is at least resolution width (for which many strong lower bounds are known), this yields strong rank lower bounds in all of the above proof systems for large classes of natural CNF formulas.

- There are strict hierarchies for $T^{cc}(k)$ and $R^{cc}(k)$ systems with respect to $k$. Specifically, for any $k$ that is $O(\log \log n)$, there are unsatisfiable CNF formulas whose proofs require large rank in $R^{cc}(k)$ but which can be refuted via polylogarithmic rank CP($k$) proofs. Rank separations between CP($k-1$) and CP($k$), between Th($k-1$) and Th($k$), and between $R^{cc}(k)$ and $T^{cc}(k+1)$ follow immediately.

- When $k$ is $O(\log \log n)$ we also derive $2^{n^{\Omega(1/k)}}$ lower bounds on the size of tree-like $T^{cc}(k)$ refutations for large classes of lifted CNF formulas. Moreover, the rank hierarchies extend to nearly exponential separations in tree-like proof size.

- A general method for producing integrality gaps for low rank $R^{cc}(2)$ inference based on related gaps for low rank resolution. This yields integrality gaps for low rank Cutting Planes and more general Th(1) inference. These gaps are optimal for MAX-$2t$-SAT.

[*]Research supported by NSF grant CCF-0830626
[†]also known as Dang-Trinh Huynh-Ngoc
[‡]Research Supported by NSERC

# 1 Introduction

Over the last decade or so there have been a large number of results proving lower bounds on the rank required to refute (or approximately optimize over) systems of constraints in a wide variety of semi-algebraic (a.k.a. polynomial threshold) proof systems. These include systems such as Lovász-Schrijver [32], Cutting Planes [19, 12], Positivstellensatz [22], Sherali-Adams [44], and Lasserre [30] proofs.

Highlights of this work include recent linear rank lower bounds for Lasserre proofs [40, 48] for many constraint optimization problems as well as rank lower bounds for semi-algebraic proof systems of varying degrees for other important optimization problems [10, 38, 26, 18, 42, 41, 17]. In addition to these rank lower bounds a few other papers also have produced superpolynomial lower bounds on the size of tree-like proofs (in which the pattern of inferences forms a tree) in specific semi-algebraic proof systems either directly [21, 20, 27] or as a consequence of the rank lower bounds [43].

Exciting and important as these results are, their proofs rely on delicate constructions of problem-specific local distributions on inputs that satisfy constraints based on the specific rules for each proof system. Furthermore, because there is not much in the way of effective reductions for such proof systems, lower bounds for one problem usually do not translate to other problems.

A very different approach for proving lower bounds for semi-algebraic proofs was developed in [2], whereby the problems of lower-bounding the rank or tree-like proof size are reduced to a lower bound problem in communication complexity. This allows the results to be applicable to a much wider class of proof systems, called $Th(k)$ proofs, which generalizes the semi-algebraic proof systems discussed above. In these systems, a proof consists of a sequence of lines, each of which is a multivariate polynomial inequality of degree at most $k$; the only requirement is that each line either expresses an input constraint or is a semantic consequence of a constant number (say two or three) of its predecessors. [2] showed that if an unsatisfiable CNF formula $G$ has a small-rank (or small tree-like size) $Th(k-1)$ refutation then, over every partition of the variables of $G$ for $k$ parties, there is an efficient $k$-party randomized NOF protocol that outputs a falsified clause in $G$. Thus to lower bound the rank of $Th(k-1)$ proofs it suffices find an unsatisfiable family of CNF formulas with the property that the $k$-player NOF complexity of this underlying search problem (outputting a falsified clause) is hard.

However, though this communication complexity approach was de-coupled from the specifics of the proof system, like the other lower bounds on semi-algebraic proofs, the reduction given in [2] was very problem-specific and delicate. One source of the difficulty was that the clause search problem needs to be hard for randomized protocols to solve but is always easy nondeterministically, as the players can easily guess and verify a violated clause. Much of the delicacy of the argument was in carefully embedding a specific candidate function (set disjointness), which appeared to have these characteristics, into the search problem of an unsatisfiable CNF.

Indeed, using a long and involved argument, [2] showed the feasibility of this communication complexity approach by constructing a particular family of CNF formulas, $(k-1)$-fold Tseitin tautologies over $\Theta(\log n)$-degree LPS expander graphs, such that lower bounds on the $k$-party randomized NOF communication complexity of the $k$-party set disjointness function yield rank and tree-like size lower bounds for $Th(k-1)$ refutations. The recent lower bounds of Lee and Shraibman [31] and Chattopadhyay and Ada [11] for the $k$-party randomized communication complexity of set disjointness thus yield unconditional rank bounds for $Th(k)$ proofs. Unfortunately, though the set disjointness bounds apply for $k$ up to $(1-o(1))\log\log n$, the details of the reduction in [2], which was claimed for each constant $k$, only apply for $k = O(\log \log \log n)$. Moreover, the method only applies to this one particular family of unsatisfiable formulas, and no other lower bounds for $Th(k)$ proofs have been known by any other method.

In this paper for the first time we provide a simple and general method that produces unsatisfiable formulas requiring proofs of large rank and tree-like size in semi-algebraic proof systems. This applies to a broad range of systems including all of $Th(k)$ for $k$ up to $(1 - o(1)) \log \log n$. Our method allows one to take any unsatisfiable formula requiring large rank in a very simple proof system, resolution, and derive new formulas that require large rank and tree-like proof size in these very powerful semi-algebraic systems. In particular, this construction applies to all formulas of large resolution width [5] since resolution width is a lower bound on resolution rank. A simplified statement of our main result is the following.

**Theorem 1.1.** *Let $F$ be any family of 3-CNF formulas in $m$ variables with resolution rank $r$. Then for any $\epsilon > 0$ and integer $k \geq 1$ there is a family of CNF formulas, $G = Lift_k(F)$ of size $n = m^{O(k)}$ such that if $k \leq (1-\epsilon) \log \log n$ then $G$ requires $Th(k+1)$ refutations of rank $r^{\Omega(1)}/\log^{O(1)} n$ and tree-size $\exp(r^{\Omega(1)})$. In particular, if $r$ is $m^{\Omega(1)}$ then $G$ requires $Th(k+1)$ rank $n^{\Omega(1/k)}$, and tree-size $\exp(n^{\Omega(1/k)})$.*

Our lower bounds are much more general than this statement. In particular, our proof shows that the lifted formula requires large rank in any proof system in which the truth of each line in a proof can be verified by an efficient $k$-party randomized communication protocol; the above theorem follows by the reduction in [2]. Our lower bounds also apply to proof systems in which individual proof lines may not be efficiently verifiable but in which any falsifying assignment at an inferred line can be traced to one of its antecedents using an efficient $k$-party randomized communication protocol.

Our method is an example of a kind of hardness amplification that we term a "hardness escalation" method, whereby one takes an object, in this case an unsatisfiable 3-CNF formula, that is hard for a weak complexity measure and produces another object, the lifted formula in this case, that is hard for a much stronger complexity measure. This is related to but different from typical hardness amplification methods where one is concerned with producing new problems for which similar classes of algorithms have much lower probability of success.

Our proof uses intuition and ideas from the pattern matrix method developed by Sherstov [46, 47] and from a related method developed earlier by Raz and McKenzie [39]. Both of these are hardness escalation methods for communication complexity. Each method begins with a computational problem that is hard for a weak complexity measure, either a relation $R$ of large decision tree complexity ([39]), or a function $f$ of large approximate polynomial degree ([46]), and extends the problem using a "pattern matrix" to produce a problem of large deterministic ([39]), or large randomized and quantum ([46]), two-party communication complexity.

We use the $k$-party generalizations of the pattern matrix method developed in [31, 11, 14, 3]. Starting with $f(\ldots, e_i, \ldots)$ on $m$ variables, and a parameter $k$, these generalizations *lift* $f$ to obtain another function $g = Lift_k(f) = f(\ldots, \psi(\cdots), \ldots)$ on $mk$ bit-strings. The transformation takes each original variable $e_i$ and replaces it by a Boolean *selector function* on $k$ bit-strings. As long as $f$ is hard in the weak measure, $g$ is hard in the $k$-player number-on-forehead (NOF) randomized communication complexity model (for a particular partition of the new variables).

A key obstacle when trying to apply the pattern matrix method to the proof complexity setting is that the approach only works with Boolean functions, and not with unsatisfiable CNF formulas.

To overcome this obstacle, we associate a family of Boolean functions $\mathcal{Z}_F$ with every unsatisfiable $F$ and show that if the hardness assumption on $F$ is satisfied then there is some function $f \in \mathcal{Z}$ that has large decision tree complexity. Furthermore, if there is an efficient communication protocol that outputs falsified clauses in $\mathrm{Lift}(F)$, then there is an efficient protocol for $\mathrm{Lift}(f)$ for any $f \in \mathcal{Z}$. In this way we are able to combine the hardness escalation ideas of [39, 46] to obtain our results.

We can also prove a converse to our result, thus completely characterizing the $\mathrm{Th}(k)$ rank of our lifted formulas. That is, in addition to deriving lower bounds on the rank of proofs for $\mathrm{Lift}_k(F)$ in terms of the resolution rank $r$ of $F$, we also show that the rank of $T^{cc}(2)$ proofs (and even resolution) proofs of $\mathrm{Lift}_k(F)$ is not much larger than $r$.

Using the above lower bounds, we are able to prove new rank separation theorems for hierarchies of polynomial threshold proof systems. By considering $\mathrm{Lift}_k(F)$ for certain unsatisfiable CNF formulas $F$ that require large rank resolution refutations but need only small rank Cutting Planes refutations, we obtain strong rank separations between the power of $T^{cc}(k+1)$ and $R^{cc}(k)$, between $\mathrm{Th}(k)$ and $\mathrm{Th}(k-1)$, and between $\mathrm{CP}(k)$ and $\mathrm{CP}(k-1)$ refutations where $\mathrm{CP}(k)$ is the natural generalization of Cutting Planes to degree $k$.

Finally, using Sherstov's strengthened degree-discrepancy lemma for 2-player communication complexity [46], we apply our techniques to prove optimal integrality gaps for a large family of optimization problems even after $n^\epsilon$ rounds of CP or $\mathrm{Th}(1)$.

**Related Work on Hardness Escalation** As mentioned above, the usual form of hardness amplification in circuit complexity is a method of amplifying the probability of error. That is, the complexity class $C$ is fixed (or nearly fixed), and the goal is to go from a function that is weakly hard (e.g., any circuit in $C$ that approximates $f$ has non-negligible probability of error) to a function that is much harder (e.g., any circuit in a slightly smaller class than $C$ that approximates $f$ has error exponentially close to $1/2$).

A different type of hardness amplification that seems to be more relevant to proof complexity is what we call *hardness escalation*. Here, we start with a function $f$ that is hard for some complexity class (where hard can be either worst case or average case), and we construct a $g$ that is hard for a larger complexity class. Hardness escalation results have been obtained in models such as in communication complexity [46], sub-exponential time complexity [23], and circuit depth ([25, 16, 39]). A similar concept called hardness condensing was introduced in [9] and some interesting results were proven for complexity classes beyond NP (with advice).

Hardness escalation for proof complexity means starting with an unsatisfiable family of formulas that is hard for some class of proof systems, and constructing a related family that is hard for a stronger class of proof systems. There have been a few papers in the proof complexity literature that have implicitly used this idea. It has been observed that if some formula $F$ requires large resolution width, then the *xorification* of $F$, obtained by replacing each variable by an xor of several variables (and then rewriting as a CNF), is hard with respect to resolution size. This idea has been used in many papers to obtain separations between various refinements of resolution, with respect to both size and space (e.g., [49, 8, 4]). More generally, [33, 35] showed how to replace variables in a somewhat hard unsatisfiable formula by hard functions in order to prove hardness escalation theorems for tree-like proof systems, with the caveat that the allowable cuts in the proof are restricted to belong to a weaker class than the hard functions. In particular, this approach fails to give lower bounds for CNF formulas. However, for certain special families of formulas, one can do better. Schoenebeck [40] has shown how to obtain rank lower bounds for Lasserre proofs based on resolution rank lower bounds for particular families of formulas,

**Outline of the Paper** Section 2 contains definitions and preliminary results that we will need. In Section 3, we prove our main result, showing how to start with an unsatisfiable CNF formula requiring large Resolution rank, and lift it to obtain another formula that is hard with respect to stronger proof systems. We present two methods of lifting $f$, one based on the tensor selector function and the second one based on the parity selector function. In Section 4 we apply our main theorems in order to prove hierarchy theorems for several proof systems. We conclude in Section 5 with a discussion and open problems.

## 2 Preliminaries

### 2.1 Functions, Search Problems and Decision Trees

For a CNF formula $F$, let clauses($F$) denote the set of clauses of $F$ and let $|F|$ denote $|\text{clauses}(F)|$. $F$ is a $t$-CNF formula iff every clause contains at most $t$ literals.

**Functional Composition** For any function $f$ on $m$ bits and any function $h$ on $s$ bits, we abuse notation slightly and use $f \circ h$ to denote the following function on $ms$ bits: $f \circ h := f(h(\cdots), \cdots, h(\cdots))$.

**Decision Problems and Search Problems** A Boolean decision problem over $n$ variables is a function from $\{0, 1\}^n$ to $\{0, 1\}$. Let $F$ be an unsatisfiable CNF formula over variables $x_1, \ldots, x_n$ consisting of $m$ clauses. The canonical Boolean relation associated with $F$ is the predicate $R_F(x, y)$, where $x$ is a vector of length $n$, and $y$ is a number, $1 \leq y \leq m$. $R_F(\alpha, \beta)$ is true if and only if $\alpha$ is a Boolean assignment and the clause $C_\beta$ in $F$ is falsified by $\alpha$. Associated with a Boolean relation $R(x, y)$ is a search problem: given $x$, output a $y$ such that $R(x, y)$ is satisfied. Given a Boolean relation $R(x, y)$, we call a function $g$ a *subfunction* of $R$ if $R(x, g(x))$ is satisfied for every $x$. In other words, $g$ is a particular function that solves the search problem associated with $R$. For example, for the canonical Boolean relation $R_F$ associated with an unsatisfiable CNF formula $F$, the search problem, $F_{\text{search}}$ is the problem of finding a violated clause given a Boolean assignment to the underlying variables of $F$. A function $g$ is a subfunction of $R_F$ if for any truth assignment $\alpha$, $g(\alpha)$ returns a clause of $F$ that is falsified by $\alpha$.

**Definition** A *decision tree* on Boolean variables $x_1, \ldots, x_n$ is a binary tree in which every non-leaf node is labeled with some $x_i$ and has two outgoing edges that are labeled with 0 and 1, and every leaf node $v$ is labeled by some value $\ell(v)$. Thus any path from the root to a leaf identifies a partial assignment to these variables. A decision tree $T$ is said to compute a function $f$ if for every leaf $v$ in $T$, its associated partial assignment determines an output for $f$ and is equal to $\ell(v)$. That is, if $\sigma$ is a partial assignment associated with a path of $T$ labelled by $\ell(v)$, then for every total assignment $\sigma'$ extending $\sigma$, $f(\sigma') = \ell(v)$. For a relation $R$, a decision tree $T$ solves the search problem associated with $R$ if it computes some subfunction of $R$. The *height* of a decision tree is the maximum length of any path from the root to a leaf. The *decision tree complexity* of $f$, denoted $D(f)$, is the minimum height of all such trees computing $f$.

## 2.2 Hard functions from hard unsatisfiable CNF formulas

Given an unsatisfiable CNF formula $F$, we will say that $F$ is *somewhat hard* if the decision tree complexity of $F_{\text{search}}$, $D(F_{\text{search}})$, is large (superpolylogarithmic in the size of $F$.) It is well known that the decision tree complexity of $F_{\text{search}}$ is equivalent to the height of any tree-like resolution refutation of $F$, or equivalently to the depth of recursion of any DPLL search procedure for $F$ [15].

Now given an unsatisfiable CNF formula $F$ that is somewhat hard, we want to identify a set $\mathcal{Z}$ of Boolean functions associated with $F$ that witnesses the hardness of $F$. Specifically, we want $\mathcal{Z}$ to have the property that if $D(F_{\text{search}})$ is large then $\mathcal{Z}$ contains a function with large decision tree complexity. This alone would be easy. However, we also want $\mathcal{Z}$ to be constructable from algorithms computing $F_{\text{search}}$.

A natural choice for the collection of functions from $F_{\text{search}}$ would be to define $f_S(\alpha) = 1$ for some $S \subseteq \text{clauses}(F)$ if and only if there is some clause in $S$ that is falsified by $\alpha$. One might hope to argue that one such $f_S$ would have decision tree complexity close to that of $F_{\text{search}}$. The obvious way to try to show this would be to reason by reduction; however, it is not clear how to construct a decision tree for $F_{\text{search}}$ from decision trees for such a collection of $f_S$ since both $f_S(\alpha)$ and $f_{\overline{S}}(\alpha)$ may equal 1. Some sort of symmetry-breaking scheme is required and this scheme must satisfy the property that for $S \subset T$ we have $f_T(\alpha) = 1$ whenever $f_S(\alpha) = 1$.

**Definition** A set $\mathcal{Z}$ of Boolean functions over the set of variables of an unsatisfiable CNF formula $F$ is said to be a *consistent system of functions for $F$* iff $\mathcal{Z} = \{f_S \mid S \subseteq \text{clauses}(F)\}$ and for any input assignment $\alpha$ there exists a clause $C$ in $F$ falsified by $\alpha$ such that for any $f_S \in \mathcal{Z}$ we have that $f_S(\alpha) = 1$ if and only if $C \in S$.

**Proposition 2.1.** *Given an unsatisfiable CNF formula $F$, any function $f^*$ that is a subfunction of $R_F$ (that is, it solves the search problem $F_{\text{search}}$) yields a consistent system $\mathcal{Z}_{f^*}$ of functions for $F$.*

*Proof.* Use the clause $C = f^*(\alpha)$ and define $f_S(\alpha) = 1$ iff $C \in S$. □

The following proposition says that any consistent system of functions for $F$ witnesses the hardness of $F$.

**Proposition 2.2.** *For any unsatisfiable CNF formula $F$ and any consistent system $\mathcal{Z}$ of functions for $F$, there exists a function $f_S \in \mathcal{Z}$ such that $D(F_{\text{search}}) \leq D(f_S) \lceil \log_2 |F| \rceil$.*

*Proof.* Build a decision tree for $F_{\text{search}}$ using binary search by querying the $f_S$ for subsets $S \subseteq \text{clauses}(F)$ to narrow down the search. The requirement of consistency ensures that the path followed by binary search on input $\alpha$ yields the falsified clause $C$. To derive the tree for $F_{\text{search}}$ replace each query of $f_S$ by the optimal decision tree for $f_S$, yielding the claimed bound. □

## 2.3 Communication Complexity

Given a function (or relation) $f$, some number $k \geq 2$ of players, and a partition of the input of $f$ for these players, communication complexity is concerned with the least amount of communication necessary between the players in order for them to compute an output of $f$. In the *number-on-the-forehead (NOF)* communication model, each player sees all inputs except the block of the partition that is assigned to him. For formal definitions, the reader is referred to [29]. In this paper, we will be only concerned with NOF randomized communication complexity.

**Definition** Let $|\mathcal{P}|$ denote the number of bits communicated in a communication protocol $\mathcal{P}$ and $\mathcal{P}(x)$ the output of the protocol on input $x$. A randomized protocol $\mathcal{P}$ is said to compute a function $f$ with error at most $\epsilon$ if on any input $x$, with probability at least $1 - \epsilon$ (over the choices of players' random coins $c$), $\mathcal{P}(x,c) = f(x)$.

If $f$ is a search problem, the standard definition (e.g. [29]) of randomized communication complexity states that $\mathcal{P}$ computes $f$ with error at most $\epsilon$ if and only if for at least $1 - \epsilon$ fraction of choices of random coins $c$, on any input $x$, $\mathcal{P}(x,c) \in f(x)$. Even for the $1 - \epsilon$ good choices of $c$, the values $\mathcal{P}(x,c)$ for different choices of $c$ may be different elements of $f(x)$. However, for our construction of hard functions from hard unsatisfiable CNF formulas, we will require a stronger notion.

**Definition** A randomized protocol $\mathcal{P}$ is said to *consistently compute* a relation $f$ with error at most $\epsilon$ if there is a function $f^*$ contained in $f$ – that is, $f^*(x) \in f(x)$ for every $x$, such that $\mathcal{P}$ computes $f^*$ with error at most $\epsilon$.

## 2.4 Proof Systems and the Complexity of Clause Search

A *proof system* for a language $\mathcal{L}$ is a polynomial time algorithm $V$ such that for all $F$, $F \in \mathcal{L}$ if and only if there exists a string $P$, referred to as a *proof*, such that $V$ accepts $(F, P)$. If $\mathcal{L}$ is the set of all unsatisfiable CNFs, or all unsatisfiable sets of inequalities, and $F \in L$, then $P$ is called a *refutation* of $F$.

A wide variety of proof systems exist in the literature. In most of these proof systems, a proof or refutation can be expressed as a sequence of *lines*, each of which is either (a translation of) an input clause or follows from some previous lines via some *inference rule*. (Inference rules that do not depend on previous lines are called *axioms*.) We call such proofs *standard proofs*. In the case that $\mathcal{L}$ is the set of unsatisfiable CNFs or propositional tautologies, in a standard proof each line represents a Boolean function on the variables of the formula, and any inference of a line representing function $g$ from lines representing functions $f_1, \ldots, f_s$ must be *sound*, in that for any Boolean assignment to their input variables $g$ must evaluate to true whenever all of $f_1, \ldots, f_s$ do. We call the maximum $s$ over all derivations in a proof its *fan-in*. A refutation of an unsatisfiable formula $f$ in a standard proof system is a sequence of formulas, where the initial formula $f$ is included as an axiom (or set of axioms), and the final formula in the sequence is the trivially false formula.

**Definition** We associate a DAG $\mathcal{G} = (V, E)$ with every standard proof $P$, where $V$ is the set of lines in $P$ and $(u, v) \in E$ if line $v$ is derived via some inference rule using line $u$. The *size* of $P$ is the number of bits in $P$, which is lower-bounded by the number of lines in $P$. The *rank* of $P$ is the length of the longest path in $\mathcal{G}$. We consider $\mathcal{G}$ to be a tree if every internal node has fanout one. (The axioms, which are not internal nodes, can be repeated.) If $\mathcal{G}$ is a tree, we say that $P$ is *tree-like*. The *size complexity* and *rank complexity* of $F$ in a standard proof system are the minimum size and minimum rank, respectively, of all proofs for $F$ in that system. Similarly, we define *tree-like size complexity* as the minimum over all proofs are restricted to be tree-like.

Note that restricting a proof to be tree-like does not increase the rank of a proof because the same line can be re-derived multiple times without affecting the rank. Tree-like size, however, can be much larger than general size.

We first mention some of the most well-studied proof systems. In each of these systems, there is a set of derivation rules (which can be thought of as inference schemas) of the form $F_1, F_2, \ldots, F_t \vdash G$ and each derivation step in a proof must be an instantiation of one of these rules.

A basic system is *resolution*, which manipulates clauses. Its only rule is the *resolution rule*: the clause $(A \vee B)$ is derived from $(A \vee x)$ and $(B \vee \neg x)$, where $A$ and $B$ are arbitrary disjunctions of literals and $x$ is a variable. A resolution refutation of an unsatisfiable CNF formula $f$ is a sequence of clauses, ending with the empty clause, such that each clause in the sequence is either a clause of $f$, or follows from two previously derived clauses via the resolution rule. The well-known connection showing that DPLL executions and tree-like resolution proofs are equivalent gives us the following proposition.

**Proposition 2.3.** *For any CNF formula $F$, the minimum rank of any resolution proof of $F$ is equal to $D(F_{\text{search}})$.*

Another proof system is the *Cutting Planes* (CP) proof system which manipulates integer linear inequalities. The two rules in the CP system are:

$$p_1 \geq 0, \ldots, p_t \geq 0 \vdash \sum_{i=1}^{t} \lambda_i p_i \geq 0, \tag{1}$$

and

$$\sum_i c a_i x_i \geq b \vdash \sum_i a_i x_i \geq \lceil b/c \rceil, \tag{2}$$

where each $p_i$ is a linear form, $a_i$, $b$, $c$, and $\lambda_i \geq 0$ are integers, and $x_i$ is a variable. In CP, we also have axioms $0 \leq x_i$ and $x_i \leq 1$, and each input clause $(\ell_1 \vee \cdots \vee \ell_t)$ is translated as $\ell'_1 + \cdots + \ell'_t \geq 1$ where $\ell' = x$ if $\ell = x$ and $\ell' = (1 - x)$ if $\ell = \neg x$. The trivial unsatisfiable formula is $0 \geq 1$. A CP refutation is a sequence of inequalities, ending with $0 \geq 1$, such that all other inequalities are either, axioms, translated input clauses, or follow from two previously derived inequalities via a CP rule.

We will consider a natural extension of CP, denoted CP($k$), in which the above CP proof rules may also be applied when the $p_i$ are allowed to be degree $k$ multivariate polynomials and the $x_i$ are replaced by degree $k$ monomials. Since the input clauses are linear there are two other rules that allows the creation of higher degree inequalities, namely:

$$p \geq 0 \vdash x_i p \geq 0$$

and

$$p \geq 0 \vdash p \geq x_i p$$

for all polynomials $p$ of degree at most $k - 1$ and variables $x_i$.

Other important well-studied proof systems are the Lovász-Schrijver proof systems (LS$_0$, LS, LS$_+$, and LS$_{+,\star}$) which manipulate polynomial inequalities of degree at most 2. These proofs use various subsets of the inference rules

$$\ell_{11} \geq 0, \ell_{12} \geq 0, \ldots \ell_{t1} \geq 0, \ell_{t2} \geq 0$$

$$\vdash \sum_{i=1}^{n} a_i(x_i^2 - x_i) + \sum_{j=1}^{t} \lambda_j \ell_{j1} \ell_{j2} + \sum_{j=t+1}^{s} \lambda_j \ell_j^2 \geq 0$$

where $\ell_j, \ell_{jb}$ are linear forms and $x_i$ are variables, and $a_i$ and $\lambda_j \geq 0$ are integers. the axioms and translations of clauses are the same as for CP. Thus, all Lovász-Schrijver proof lines are degree 2 polynomial inequalities. On can generalize these proof rules to $LS^k_{+,\star}$ proofs in which one is allowed to multiply arbitrary inequalities, use $x_i^2 = x_i$ and add the squares of higher degree terms, provided that each quantity in the inference is syntactically a polynomial of degree at most $k$.

Each of the above proof systems has a specific set of inference rule schemas, which allows them to have polynomial-time verifiers. We also consider more powerful *semantic proof systems* which restrict the form of the lines and the fan-in of the inferences but dispense with the requirement of a polynomial-time verifier and allow any semantically sound inference rule with a given fan-in. (Each line is a clause or follows via some semantic inference rule.) The fan-in must be restricted because the semantic rules are so strong. The following strong semantic proof system was introduced in [2].

**Definition** For integer $k \geq 1$, we denote by Th($k$) the semantic proof system whose proofs have fan-in 2, each line is a polynomial inequality of degree at most $k$, and input clauses and axioms are represented as linear inequalities as in the definition of CP above.

Without loss of generality via Caratheodory's Theorem, for formulas in $n$ variables, in the case of CP the fan-in of inferences is at most $n$ and in the cases of LS$_0$, LS, and LS$_+$, the fan-in of inferences is at most $(n+1)^2$. From this we immediately obtain the following:

**Proposition 2.4.** *(1) Any CP proof of size (tree-like size) $S$ and rank $r$ can be converted to a* Th(1) *proof of size (tree-like size) $O(S)$ and rank $O(r \log n)$. (2) Any LS$_0$, LS, or LS$_+$ proof of size (tree-like size) $S$ and rank $r$ can be converted to a* Th(2) *proof of size (tree-like size) $O(S)$ and rank $O(r \log n)$.*

Moreover, it is not hard to show that one can extend the above simulations by Th($k$) proofs to CP($k$) and $LS^k_{+,\star}$.

The Sherali-Adams and Lasserre proof systems introduce new variables for subsets of input variables of bounded size (which is called the rank of such proofs). Monomials of degree $k$ represent the intended meaning of these added variables so Th($k$) proofs of rank $k$ also efficiently simulate rank $k$ Sherali-Adams proofs and rank $k/2$ Lasserre proofs.

In this paper we also consider more general semantic proof systems even than Th($k$), namely those for which the fan-in is bounded and the truth value of each line can be computed by a multiparty communication protocol.

**Definition** For any $k, C \geq 1$, we denote by $T^{cc}(k, C)$ the semantic proof system of fan-in 2 in which each proof line is a Boolean function whose value, for every partition of the input variables into $k$ groups[1], can be computed by a $C$-bit randomized $k$-party NOF communication protocol of error at most $1/3$. Both $k$ and $C$ may be integer functions of the input size of the formulas. In keeping with the usual notions of what constitutes efficient communication protocols, we use $T^{cc}(k)$ to denote the union of all $T^{cc}(k, C)$ over all $C$ in $\log^{O(1)} n$.

Note that via standard boosting, we can replace the error $1/3$ in the above definition by $\epsilon$ at the cost of increasing $C$ by an $O(\log 1/\epsilon)$ factor. Therefore, without loss of generality, in the definition of $T^{cc}(k)$ we can assume that the error is at most $1/n^{\log^{\Omega(1)} n}$.

---

[1] We note that one could alternatively define $T^{cc}(k, C)$ systems based on a *fixed* partition of the inputs. While this definition might yield a stronger proof system, it would complicate the notation without changing our results in any significant way.

Note also that a semantic proof of rank $r$ that satisfies the same conditions as a $T^{cc}(k,C)$ proof except that it has rules of fan-in at most $t \geq 2$ can be simulated by a $T^{cc}(k, 2Ct\log_2 t)$ proof of rank $r\log_2 t$ by replacing each inference by a binary tree of height $\log_2 t$ in which lines of internal nodes are conjunctions of their predecessors.

For polylogarithmic $k$, the following lemma shows that $\text{Th}(k)$ is a subclass of $T^{cc}(k+1)$.

**Lemma 2.5.** *For some constant $c > 0$, every $\text{Th}(k)$ refutation of a CNF formula on $n$ variables is a $T^{cc}(k+1, ck^3 \log^2 n)$ proof.*

*Proof.* By the well-known result of Muroga [34], linear threshold functions on $n$ Boolean values only require coefficients of $O(n \log n)$ bits. Since a degree $k$ threshold polynomial is a linear function on at most $n^k$ monomials, it is equivalent to a degree $k$ threshold polynomial with coefficients of $O(kn^k \log n)$ bits. As shown in [2], over any input partition there is a randomized $(k+1)$-party communication protocol of cost $O(k \log^2 s)$ and error $\leq 1/s^{\Omega(1)}$ to verify a degree $k$ polynomial inequality with $s$-bit coefficients. $\square$

We also define another class of proofs based on $k$-party communication complexity that we will see is even more general than $T^{cc}(k,C)$.

**Definition** For any integer functions $k, C \geq 1$, we denote by $R^{cc}(k,C)$ the semantic proof system of arbitrary fan-in in which each proof line is a Boolean function such that the proof satisfies the following property: for every partition of the input variables into $k$ groups, and every inference of $B$ from $A_1, \ldots, A_s$ in the proof, there is a $C$-bit randomized $k$-party NOF communication protocol of error at most $1/3$ that computes a (partial) function $f_{A_1,\ldots,A_s \vdash B}$ from the inputs to the set $[s]$ such that on every input $\alpha$, if $B$ evaluates to false on input $\alpha$ then $A_{f_{A_1,\ldots,A_s \vdash B}(\alpha)}$ evaluates to false on input $\alpha$.

We write $R^{cc}(k)$ to denote the union of all $R^{cc}(k,C)$ over all $C$ in $\log^{O(1)} n$.

The following is immediate:

**Lemma 2.6.** *Every $T^{cc}(k,C)$ proof is an $R^{cc}(k,C)$ proof.*

*Proof.* The inferences in the $T^{cc}(k,C)$ are all of fan-in at most 2 and hence derive each line $B$ from some lines $A_1$ and $A_2$. To compute the function $f_{A_1, A_2 \vdash B}$ the players evaluate $A_1$ on input $\alpha$ using the protocol given by the $T^{cc}(k,C)$ proof. If that evaluates to false then they output 1; otherwise, they output 2. $\square$

We can sharpen this relationship further. The following is a standard method for strengthening a proof system $S$ by adding resolution rules over the lines of $S$ [28].

**Definition** Given a proof system $S$, we define related proof system $R(S)$ as follows: Lines of $R(S)$ are unordered disjunctions of lines of $S$ and their negations. For every inference rule in $S$, $A_1, \ldots, A_t \vdash B$, there is the corresponding rule $(G \vee A_1), \ldots, (G \vee A_t) \vdash (G \vee B)$ where $G$ is an arbitrary disjunction of lines of $S$ and their negations. In addition there are extended resolution rules that allow the introduction of new disjuncts, $G \vdash (G \vee A_1 \vee \ldots \vee A_t)$, or cuts on lines of $S$, namely $(G \vee A), (H \vee \neg A) \vdash (G \vee H)$, where $A$ is a line of $S$ and $G$ and $H$ are arbitrary disjunctions of lines of $S$ and their negations.

**Lemma 2.7.** *Every $R(T^{cc}(k,C))$ proof is an $R^{cc}(k,C)$ proof.*

*Proof.* For rules that correspond to rules of $T^{cc}(k,C)$ we apply the simple argument from Lemma 2.6 on the lines that are not common to all formulas. For the resolution rules, observe that the players only need to evaluate the line $A$ to determine whether to select $(G \vee A)$ or $(H \vee \neg A)$. □

In particular, this shows via Lemma 2.5 that $R^{cc}(k+1, ck^3 \log^2 n)$ proofs include the proof system $R(\text{Th}(k))$ (suggested by Hirsch). It is not clear whether one can efficiently simulate $R(\text{Th}(k))$ using $T^{cc}(k)$ proofs.

The following lemma, which is implicit in [2], gives the key relationships between $T^{cc}(k)$ and $R^{cc}(k)$ proofs and randomized communication protocols that consistently compute $F_{\text{search}}$.

**Lemma 2.8.** *Let $F$ be a CNF formula in $n$ variables and $\epsilon > 0$.*

(i) *If $F$ has an $R^{cc}(k, C)$ refutation of rank $r$ then, over every partition of the variables, there is an $\epsilon$-error randomized $k$-party communication protocol $\mathcal{P}$ consistently computing $F_{\text{search}}$ such that $|\mathcal{P}|$ is $O(Cr \log(r/\epsilon))$.*

(ii) *If $F$ has a tree-like $T^{cc}(k, C)$ refutation of size $S$. then, over every partition of the variables, there is an $\epsilon$-error randomized $k$-party communication protocol $\mathcal{P}$ consistently computing $F_{\text{search}}$ such that $|\mathcal{P}|$ is $O(C \log S \log(\log S/\epsilon))$.*

*Proof.* First assume that we have a rank $r$ refutation in $R^{cc}(k,C)$. On input $\alpha$, the $k$ players backtrack from the last derived inequality in the proof ($0 \geq 1$) to find some clause that is falsified by $\alpha$. When they are at a line $B$ that follows from lines $A_1, \ldots, A_s$ in the proof, they run the protocol for $f_{A_1,\ldots,A_s \vdash B}$, implied by the $R^{cc}(k,C)$ definition for the inference at $B$, $O(\log(r/\epsilon))$ times and take the majority answer to reduce its error below $\epsilon/r$. Then the players move to the line indicated by that answer. The probability that this protocol makes an error is at most the sum of all error probabilities on any path in the proof. Since the last line evaluates to false on input $\alpha$, in the case that there is no error the players will return a fixed clause in the proof that is falsified by $\alpha$, which implies that they consistently compute $F_{\text{search}}$.

For the second case of a size $S$ tree-like refutation, there is some line in the refutation that is derived from between $S/3$ and $2S/3$ of the lines of the refutation tree. The players first evaluate that line with error at most $\epsilon/(2\log_2 S)$ by repeating the protocol $O(\log(\log S/\epsilon))$ times. If the line evaluates to false then they continue within that subtree; otherwise, they remove the nodes of that subtree. This is done recursively until a falsified clause is found. The depth of recursion is at most $2\log_2 S$. The rest is similar to the first case. □

## 3 Hardness Escalation for CNF formulas

This section proves our results on hardness escalation. The high level idea is as follows. Recall that an unsatisfiable $t$-CNF formula $F$ is somewhat hard if $F_{\text{search}}$ requires a large height decision tree. Starting with a somewhat hard unsatisfiable $t$-CNF formula $F$ over variables $e_1, \ldots, e_m$, we build a new CNF formula $G = \text{Lift}(F)$ of size $m^{O(t)}$ by lifting $F$ based on some function $\psi$ that encodes $e_i$ using a larger collection of input bits. This lifting over CNF is adapted from previous work for Boolean functions, which we review next.

## 3.1 Lifting Decision Tree Complexity to $k$-Party Communication Complexity

In this section, we show how to lift a Boolean function $f$ to obtain another function $g := f \circ \psi_k$ for some Boolean function $\psi_k$, where $\psi_k$ can be thought of as a simple encoding of a variable of $f$ using some number of new variables.

**Definition** Given $k, s > 0$ and a domain $A$, a function $\psi_k : \{0,1\}^s \times A^k \mapsto \{0,1\}$ is called a *selector* if there is some $h : A^k \mapsto [s]$ such that $\psi_k(x, y_1, \ldots, y_k) = x_{h(y_1, \ldots, y_k)}$ for every $x \in \{0,1\}^s$ and $y_i \in A$. Informally, $\psi_k$ outputs a bit in $x$ that is selected by the values of $y_1, \ldots, y_k$.

There are two specific selector encodings $\psi_k$ that we are interested in: the tensor selector $\psi_{k,\ell}^{\mathrm{T}}$ and the parity selector $\psi_{k,a}^{\oplus}$. In the tensor selector $\psi_{k,\ell}^{\mathrm{T}}(x,y)$, we have $s = \ell^k$ and $A = [\ell]$, and we think of $x \in \{0,1\}^s$ as indexed by $A^k$ and hence $h(\cdot)$ is just the identity function on $A^k$. In the parity selector $\psi_{k,a}^{\oplus}(x,y)$, we have $s = 2^a$, $A = \{0,1\}^a$, and we think of $x$ as indexed by $a$-bit arrays and $h(y_1, \ldots, y_k) = y_1 \oplus \cdots \oplus y_k$.

Given our initial function $f$ over variables $x$, we define $g$, the $(k+1)$-lifted version of $f$, to be the function $f \circ \psi_k$.

It is not hard to see that if the decision tree complexity of $f$ is $d$, then for any $k \geq 2$, and over any partition of the variables into $k$ groups, there is a $k$-party communication protocol computing $g$ of cost approximately $d \cdot c$, where $c$ is the cost of computing $\psi_k$. The $k$ players just simulate the decision tree for $f$ and the cost of computing any single variable in $f$ encoded by $\psi_k$ is $c$ bits. If $\psi_k$ is simple enough, and therefore $c$ is negligible, then this cost is approximately equal to $d$. Ideally, we would like to argue that this is the best that the players can do. Intuitively, since we have encoded each input bit in $f$ indirectly, the players need to communicate $\Omega(1)$ bits in order to be able to "learn" any single bit. If the decision tree complexity of $f$ is large, we would hope that $g$ has large communication complexity. Recent results in communication complexity show that we cannot do much better than the above trivial protocol, subject to some constraints on $\psi_k$.

We need the following approximation notion to bridge decision tree complexity and communication complexity; this notion of approximating a real-valued function is polynomially related to decision tree complexity.

**Definition** Given any $0 \leq \epsilon < 1$, the $\epsilon$-*degree* of a real-valued function $f$, $\deg_\epsilon(f)$, is the smallest $d$ for which there exists a multivariate real-valued polynomial $p$ of degree $d$ such that $||f - p||_\infty = \max_x |f(x) - p(x)| \leq \epsilon$.

**Proposition 3.1** ([36, 1]). *For every Boolean function $f$, $\deg_{5/6}(f) \leq D(f) \leq (4 \deg_{5/6}(f))^6$.*

Finally we state the communication lower bounds for $g = f \circ \psi_k$. The following input partition is always assumed when the communication complexity of $g$ is discussed: there are $k+1$ players and for each input $(x, y_1, \ldots, y_k)$ to each $\psi_k$, player 0 is assigned $x$, and each player $i$, for $1 \leq i \leq k$, is assigned $y_i$. Intuitively, the inputs $y_1, \ldots, y_k$ given to players 1 through $k$ determine which bits of $x$ (player 0's input) are given to $f$. The next two results say that, when $\psi_k$ is either $\psi_{k,\ell}^{\mathrm{T}}$ or $\psi_{k,a}^{\oplus}$, and the encoding $\psi_k$ is over a large enough number of new variables, then the communication complexity of $g$ is polynomial related to $D(f)$ (up to a factor depending only on $k$).

**Theorem 3.2** ([11]). *Let $f : \{0,1\}^m \mapsto \{0,1\}$ with $5/6$-degree $d > 2$. If $\ell > \frac{2^{2^{k+1}} kem}{d}$ then any $(k+1)$-party communication protocol $\mathcal{P}$ computing $g = f \circ \psi_{k,\ell}^{\mathrm{T}}$ with error $1/3$ must have $|\mathcal{P}| = \Omega(\frac{d}{2^k})$.*

**Theorem 3.3** ([3]). *Let $f : \{0,1\}^m \mapsto \{0,1\}$ with 5/6-degree $d > 2$. If $2^a > \frac{2^{2^{k+1}+2k}em}{d}$ then any $(k+1)$-party communication protocol $\mathcal{P}$ computing $g = f \circ \psi_{k,a}^{\oplus}$ with error $1/3$ must have $|\mathcal{P}| = \Omega(\frac{d}{2^k})$.*

The first theorem uses the tensor selector while the second uses the parity selector. We will use both of them to prove lower bounds for $T^{cc}(k)$ and $R^{cc}(k)$ proof systems. The parity selector has an advantage that it needs fewer bits to encode each variable. Thus, it will give stronger proof complexity lower bounds as a function of the number of variables of the formula (though it is no more efficient with respect to formula size). In contrast, the advantage of the tensor selector is that a CNF formula that is lifted based on the tensor selector is easier to refute by small degree threshold proof systems so we will be able to use this selector to prove rank separations for the hierarchies of $T^{cc}(k)$ and $R^{cc}(k)$ proof systems.

**Overview of the Hardness Escalation Argument**

Before giving the formal construction and proofs for the two selectors, we present a brief overview of our argument. Let $F$ be any $t$-CNF over the variables $e_1, \ldots, e_m$. We want to describe how to lift $F$ to obtain another unsatisfiable formula $G$, where now $G$ is harder than $F$. Every variable $e_i$ of $F$ will be replaced by a set of variables $V_i$. The $V_i$ variables be comprised of $k+1$ sets of variables: $x$, and $y_1, \ldots, y_k$. As in the previous section, there will be a selector function $\psi_k$ which will use the $y$ variables to select one $x$ variable to represent $e_i$. The clauses in $G$ will state that the $V_i$ variables represent a valid $\psi$-encoding, and that with respect to this encoding, $F$ is true. We want to show that $G$ is even harder than $F$, i.e., that $G$ requires large $T^{cc}(k)$ rank. By Lemma 2.8, we know that if $G$ has low $T^{cc}(k)$ rank, then there is an efficient $k$-party protocol for solving the search problem associated with $G$, $G_{\text{search}}$. Thus to prove a $T^{cc}(k)$ rank lower bound for $G$, it suffices to prove that $G_{\text{search}}$ is hard in the $k$-party NOF model.

Now any function associated with $G$ is also a lifting of the corresponding function associated with $F$. In particular, $G_{\text{search}} = F_{\text{search}} \circ \psi_k$. The intuition for why it should be hard is similar to that of the lifting of Boolean functions: here $G_{\text{search}}$ is a lifting of $F_{\text{search}}$, and the decision tree complexity of $F_{\text{search}}$ is large. To prove this, assume for sake of contradiction that $G_{\text{search}}$ is easy for $k$-party communication. Then the $k$ players can efficiently compute $G_{\text{search}}$ over the variables $V_i$. This in turn means that given the variables $V_i$, they can efficiently compute $F_{\text{search}}(e_1, \ldots, e_m)$, where each $e_i = \psi_k(V_i)$. It follows that there exists a consistent system $\mathcal{Z}$ of functions for $F$ such that for any function $f_S \in \mathcal{Z}$, the players can easily compute $f_S \circ \psi_k$. In other words, the lifting of any $f_S \in \mathcal{Z}$ is easy for $k$-party communication. It then follows that for appropriate choices of $\psi_k$, any function in $\mathcal{Z}$ has low decision tree complexity. Then by Proposition 2.2, we can conclude that the decision tree complexity of $F_{\text{search}}$ is small, contradicting our assumption. We now proceed to the formal arguments for each of the selector functions.

## 3.2 Hardness Escalation Based on the Tensor Selector

Let $F$ be any $t$-CNF over the variables $e_1, \ldots, e_m$. Parametrized by $k, \ell \geq 2$, $G = \text{Lift}_{k,\ell}^T(F)$ is a CNF formula defined over $m$ sets of variables $V_1, \ldots, V_m$, where each $V_i$ is further partitioned into two sets $X_i$ of size $\ell^k$ and $Y_i$ of size $k\ell$. Intuitively, every $V_i$ is an encoding of $e_i$ based on $\psi_{k,\ell}^T$. Each $X_i$ represents a $k$-dimensional tensor of size $\ell^k$ each of whose cells $c$ is associated with a variable $x_{i,c} \in X_i$. $Y_i$, which is indexed as $\{y_{i,p,a} : 1 \leq p \leq k, 1 \leq a \leq \ell\}$, selects a unique cell $c$ in this tensor as follows: For each $p \in [k]$, exactly one of the variables $y_{i,p,a}$ for $a \in [\ell]$ is true, and the value $a_p$ such that $y_{i,p,a_p}$ is true is the $p$-th coordinate of $c$. Every clause in $F$ is then transformed into a set of clauses over these $V_i$. Formally, the clauses in $G$ consist of:

- For $1 \leq i \leq m, 1 \leq j \leq k$, exactly one of $y_{i,p,1}, \ldots, y_{i,p,\ell}$ is 1:

  (I) $y_{i,p,1} \vee \cdots \vee y_{i,p,\ell}$

  (II) $(1 \leq a < a' \leq \ell)$: $\neg y_{i,p,a} \vee \neg y_{i,p,a'}$

- For every clause, say $\neg e_{i_1} \vee e_{i_2} \vee \cdots \vee e_{i_t}$, in $F$ and for every $t$-tuple of cells $(c_1, \ldots, c_t)$, if $Y_{i_1}$ selects $c_1$, $Y_{i_2}$ selects $c_2$, etc., then $\neg x_{i_1,c_1} \vee x_{i_2,c_2} \cdots \vee x_{i_t,c_t}$ must be satisfied: this is translated into one clause of $tk + t$ literals. For example, if the coordinates of $c_1, \ldots, c_t$ are $(a_1^1, \ldots, a_k^1), \ldots, (a_1^t, \ldots, a_k^t)$, respectively, then the clause would be:

  (III) $\neg y_{i_1,1,a_1^1} \vee \cdots \vee \neg y_{i_1,k,a_k^1} \vee \cdots \vee \neg y_{i_t,1,a_1^t} \vee \cdots \vee \neg y_{i_t,k,a_k^t} \vee \neg x_{i_1,c_1} \vee x_{i_2,c_2} \vee \cdots \vee x_{i_t,c_t}$

The next proposition shows that as long as the clauses of $F$ are not too large, then $G$ is also not too large, and that $G$ is unsatisfiable as long as $F$ is.

**Proposition 3.4.** *If $F$ is a $t$-CNF over $m$ variables, then $G = \text{Lift}_{k,\ell}^T(F)$ is a CNF formula of $|F|\ell^{tk} + O(mk\ell^2)$ clauses of size at most $\max\{tk+t, \ell\}$ over $n = m(\ell^k + k\ell)$ variables. Furthermore, if $F$ is unsatisfiable, then so is $G$.*

We say that an assignment to $X_1, Y_1, \ldots, X_m, Y_m$ of $G$ is a *valid encoding* of an assignment to variables $e_1, \ldots, e_m$ of $F$ if all clauses (I) and (II) are satisfied and for every $i$, $x_{i,c} = e_i$ where $c$ is selected by $Y_i$.

We fix the following input partition to $k+1$ players when discussing the communication complexity of $G_{\text{search}}$: player 0 is assigned all of the $X_i$'s, and each player $p$, for $1 \leq p \leq k$, is assigned $\{y_{i,p,a} : 1 \leq i \leq m, 1 \leq a \leq \ell\}$.

The following lemma says that if $G_{\text{search}}$ is easy in communication complexity, then there exists a consistent system $\mathcal{Z} = \{f_S : S \subseteq \text{clauses}(F)\}$ for $F$ such that for every $f_S \in \mathcal{Z}$, computing $f_S \circ \psi_{k,\ell}^T$ is also easy in communication complexity.

**Lemma 3.5.** *Given any unsatisfiable $t$-CNF formula $F$ and $G = \text{Lift}_{k,\ell}^T(F)$. Suppose that there is a $(k+1)$-party communication protocol $\mathcal{P}$ consistently computing $G_{\text{search}}$ with error $\epsilon$ such that $|\mathcal{P}| \leq C$. Then there exists a consistent system $\mathcal{Z} = \{f_S : S \subseteq \text{clauses}(F)\}$ of functions for $F$ such that for every $S$, there is a $(k+1)$-party communication protocol $\mathcal{P}_S$ consistently computing $f_S \circ \psi_{k,\ell}^T$ with error $\epsilon$ such that $|\mathcal{P}_S| \leq C$.*

*Proof.* For every input assignment $\alpha$ to $F$, we fix any input assignment $\alpha^T$ to $G$ that is a valid encoding of $\alpha$. Let $g^*$ be the subfunction of $G_{\text{search}}$ that is computed by $\mathcal{P}$.

We first observe that on any input assignment $\alpha$ and $\alpha^T$, $g^*(\alpha^T)$ always outputs a type (III)-clause. This is because $\alpha^T$ is a valid encoding. This clause corresponds to a unique clause in $F$ that is falsified by $\alpha$. Thus $g^*$ uniquely determines a subfunction $f^*$ of $F_{\text{search}}$.

Given $f^*$, we define the consistent system $\mathcal{Z} = \mathcal{Z}_{f^*}$ for $F$ using the construction in Proposition 2.1. For every $f_S \in \mathcal{Z}$, the protocol $\mathcal{P}_S$ for $f_S \circ \psi_{k,\ell}^T$ is adapted from $\mathcal{P}$ in the straightforward way. □

The next theorem is our main result on the proof complexity of $G$ which glues all the parts together.

**Theorem 3.6.** *There are absolute constants $c, c' > 0$ such that the following holds. Let $F$ be any $t$-CNF formula on $m$ variables having resolution rank at least $r$ and let $G = \mathrm{Lift}_{k,\ell}^{\mathrm{T}}(F)$ for $\ell \geq \frac{c2^{2^{k+1}} km}{(r/\log|F|)^{1/6}}$. Then for any $C$ and $M = c'(r/\log_2|F|)^{1/6}/(C2^k)$,*

- *any $R^{cc}(k+1, C)$ refutation of $G$ of rank $R$ must have $R \log_2 R \geq M$, and*
- *any tree-like $T^{cc}(k+1, C)$ refutation of $G$ of size $S$ must have $\log S \log \log S \geq M$.*

*Proof.* We will prove only the first part, with the second part follows similarly.

Let $P$ be a $R^{cc}(k+1, C)$ refutation of $G$ of rank $R$. Lemma 2.8, there exists a $(k+1)$-party protocol $\mathcal{P}$ consistently computing $G_{\text{search}}$ of error $1/3$ such that $|\mathcal{P}|$ is $O(CR \log R)$.

Now on the one hand, by Lemma 3.5, there exists a consistent system $\mathcal{Z} = \{f_S : S \subseteq \text{clauses}(F)\}$ of functions for $F$ such that for every $S$, there exists a $(k+1)$-party protocol $\mathcal{P}_S$ computing $f_S \circ \psi_{k,\ell}^{\mathrm{T}}$ of error $1/3$ such that $|\mathcal{P}_S|$ is $O(CR \log R)$.

On the other hand, by Proposition 2.3, the assumption on the resolution rank of $F$ implies that $D(F_{\text{search}}) \geq r$. By Proposition 2.2, there exists a function $f_S \in \mathcal{Z}$ such that

$$D(f_S) \geq \frac{D(F_{\text{search}})}{\lceil \log_2 |F| \rceil} = \frac{r}{\lceil \log_2 |F| \rceil}.$$

By Proposition 3.1, we have $d = deg_{5/6}(f_S) \geq (D(f_S))^{1/6}/4 \geq (\frac{r}{\log_2 |F|})^{1/6}/4$.

Finally, by Theorem 3.2, we must have $CR \log R$ that is $\Omega(d/2^k)$ which is $\Omega((r/\log_2 |F|)^{1/6}/2^k)$. $\square$

We note that we have a somewhat matching upper bound on the rank complexity of $G$.

**Lemma 3.7.** *Let $F$ be a $t$-CNF formula on $m$ variables having resolution rank $r$. There is some absolute constant $c > 0$ such that for any $\ell \geq 1$, there is an $T^{cc}(2, \log_2(rk\ell))$ proof of $G = \mathrm{Lift}_{k,\ell}^{\mathrm{T}}(F)$ of rank at most $crk \log_2 \ell$.*

*Proof.* The main idea is to first build a decision tree for $G_{\text{search}}$ using the decision tree for $F_{\text{search}}$. The key idea in the search is that for every $e_i$ there is precisely one variable $x_{i,(a_1,\ldots,a_k)}$ for $a_1, \ldots, a_k \in [\ell]$ whose value will replace that of $e_i$ in evaluating $G$. This selection is determined by the one tuple for which all of $y_{i,1,a_1}, \ldots, y_{i,k,a_k}$ evaluate to 1.

Whenever the decision tree for $F_{\text{search}}$ queries a variable $e_i$, the decision tree over linear inequalities for $G_{\text{search}}$ does $k$ binary searches where the $p$-th one queries inequalities of the form $\sum_{a \in [j,j']} y_{i,p,a} \geq 1$ to find the unique $a_p$ such that $y_{i,p,a_p} = 1$. At the leaf of this search corresponding to the tuple $(a_1, \ldots, a_k)$, the query of $e_i$ is replaced by a query to $x_{i,(a_1,\ldots,a_k)}$.

It remains to convert this decision tree to an $T^{cc}(2)$ refutation. We follow the standard conversion of decision trees to proofs implicit in the equivalence in Proposition 2.3. Each node of the decision tree for $G_{\text{search}}$ will be a line in the new proof. Each such node $v'$ is associated with a node $v$ of the derivation for $F_{\text{search}}$ which also corresponds to a line in the resolution refutation of $F$ that is some clause $C_v$ on the $e_i$ variables. The line labelling $v'$ in the proof of $G$ will consist of a disjunction of several literals and one polynomial inequality associated with the current binary search. In particular, for each literal $e_i^b$ in $C_v$, if the branch on which $v'$ lies has determined that $x_{i,(a_1,\ldots,a_k)}$ is the selected literal to replace $e_i$ then the disjunction will include $\neg y_{i,1,a_1} \vee \ldots \vee \neg y_{i,k,a_k} \neg \vee x^b_{i,(a_1,\ldots,a_k)}$ whose negation indicates the selection and the value substituted for $e_i$. However, at node $v'$, only part of the next level of search may be completed. Suppose that $p - 1 < k$ binary searches are completed at $v'$ for the current branch variable $e_{i'}$. In this case we add $\neg y_{i',1,a_1} \vee \ldots \vee \neg y_{i',p-1,a_{p-1}}$ to the disjunction at $v'$. Finally, if the current binary search has been restricted to a range $[j, j']$ then we add one more disjunct: the linear inequality $\sum_{a \in [j,j']} y_{i',p,a} \leq 0$. (If $j = j'$ this is equivalent to $\neg y_{i',p,j}$.) (Note that in moving down the proof tree, as we start a binary search we have no such linear inequality disjunct but we can add $\sum_{a \in [\ell]} y_{i',p,a} \leq 0$ via the axiom on the selector.)

It is clear that the proof tree is binary and each line can be viewed as a disjunction of two linear inequalities on at most $rk\ell$ binary values which can be evaluated efficiently by a 2-party randomized protocol. □

**Corollary 3.8.** *Let $t$ be some constant. Suppose that a family of polynomial-size $t$-CNF formulas $F$ on $m$ variables has resolution rank complexity $r = r(m)$ that is $m^{\Omega(1)}$. Then, for every constant $\epsilon > 0$ and $k \geq 1$, there is a family of CNF formulas $G = \mathrm{Lift}_k^{\mathrm{T}}(F)$ on $n$ variables of size $n^{O(t)}$ such that if $k \leq (1 - \epsilon) \log \log n$ then*

- *$G$ requires $R^{cc}(k+1)$ refutation rank complexity $\Omega(r^{1/7}) = n^{\Omega(1/k)}$;*
- *there is a $T^{cc}(2)$ refutation of $G$ of rank $O(r \log n)$;*
- *$G$ requires $T^{cc}(k+1)$ tree-size $\exp(n^{\Omega(1/k)})$.*

*Proof.* Apply Theorem 3.6 with $\ell$ being the least integer satisfying the constraint; that is, $\ell = \frac{c'' 2^{2^{k+1}} km}{(r/\log m)^{1/6}}$ for some constant $c'' > 0$ since $|F|$ is polynomial in $m$.

By Proposition 3.4, the resulting formula $G = \mathrm{Lift}_k^{\mathrm{T}}(F)$ has $n = m(\ell^k + k\ell)$ variables. Now for $k \leq (1 - \epsilon) \log \log n$ and since $r$ is $m^{\Omega(1)}$, for sufficiently large $n$ we have $\ell^k + k\ell < 2^{2^{k+1}k}(c''m)^k \leq n^{1-\delta}(c''m)^k$ for some constant $\delta > 0$. It follows that $n \leq n^{1-\delta} m(c''m)^k$ and hence $m$ is $n^{\Omega(1/k)}$. Also by Proposition 3.4, $|G|$ is $O(|F|\ell^{tk} + mk\ell^2)$, which is $n^{O(t)}$.

Suppose that there is a $T^{cc}(k+1)$ refutation $P$ of $G$ of rank $R$. Hence by definition, there is some constant $\beta > 0$ such that $P$ is a $T^{cc}(k+1, \log^\beta n, 1/3)$ refutation. By Theorem 3.6, we have that $R \log R$ is $\Omega((r/\log m)^{1/6}/(2^k \log^\beta n))$. Thus for sufficiently large $n$, $R$ is $\Omega(r^{1/7}) = n^{\Omega(1/k)}$ since $r$ is $m^{\Omega(1)}$.

The rank upper bound follows easily from Lemma 3.7 and the proof for the tree-like size lower bound is similar. □

In particular, by Proposition 2.5, Corollary 3.8 applies to all $\mathrm{Th}(k)$ proof systems.

### 3.3 Hardness Escalation Based on the Parity Selector

Let $F$ be any $t$-CNF over the variables $e_1, \ldots, e_m$. Parametrized by $k, a \geq 2$, $\mathrm{Lift}_{k,a}^\oplus(F)$ is a CNF defined over $m$ sets of variables $V_1, \ldots, V_m$, where each $V_i$ is further partitioned into two sets $X_i$ and $Y_i$. The difference here with $\mathrm{Lift}_{k,a}^\oplus(F)$ is that every $V_i$ is an encoding of $e_i$ based on $\psi_{k,a}^\oplus$. That is, each $X_i$ has $2^a$ variables that are indexed by $a$-bit vectors, each $Y_i = \{y_{i,p,b} : 1 \leq p \leq k, 1 \leq b \leq a\}$ has $ka$ variables, and each $Y_i$ selects a unique $a$-bit vector $c$ with $c_b = \oplus_{p=1}^k y_{i,p,b}$, for $1 \leq b \leq a$. The clauses in $\mathrm{Lift}_{k,a}^\oplus(F)$ consist of:

(*) For every clause, say $e_{i_1} \vee \cdots \vee e_{i_t}$, in $F$ and for every $t$-tuple of $a$-bit vectors $(c_1, \ldots, c_t)$, if $Y_{i_1}$ selects $c_1$, $Y_{i_2}$ selects $c_2$, etc., then $x_{i_1,c_1} \vee \cdots \vee x_{i_t,c_t}$ must be satisfied. For every clause and $t$-tuple, this is translated into $\leq 2^{tka}$ clauses of size $tka + t$ in the straightforward way. That is, there are $\leq 2^{ka}$ assignments to the bits in $Y_{i_1}$ that make them select $c_1$, and similarly for $Y_{i_2}$, etc. There is one clause, similar to the clauses of type (III) in the tensor selector case, corresponding to each such assignment.

**Proposition 3.9.** *If $F$ is a $t$-CNF over $m$ variables, then $G = \text{Lift}_{k,a}^{\oplus}(F)$ is a CNF formula of at most $|F|2^{tka+ta}$ clauses of size at most $ka + t$ over $n = m(2^a + ka)$ variables. Furthermore, if $F$ is unsatisfiable, then so is $G$.*

The rest of the proofs for this section are very similar to those in the last section. The first difference is that since $\psi_{k,a}^{\oplus}$ gives a more efficient encoding than $\psi_{k,\ell}^{\text{T}}$, the blow-up in the number of variables of $G$ is significantly reduced. The second difference is that, here, $G$ has a small rank resolution refutation, as opposed to a $CC(2)$ refutation in the last section when the lifting was done using tensor-encoding. There, small rank resolution refutation was impossible because the final clauses were too large.

The proof of the following theorem, which lower bounds the proof complexity of $\text{Lift}_{k,a}^{\oplus}(F)$ in terms of that of $F$, is identical to that of Theorem 3.6 except that Theorem 3.3 for $\psi_{k,a}^{\oplus}$ is used in place of Theorem 3.2.

**Theorem 3.10.** *There are absolute constants $c, c' > 0$ such that the following holds. Let $F$ be any $t$-CNF formula on $m$ variables having resolution rank at least $r$ and let $G = \text{Lift}_{k,a}^{\oplus}(F)$ for $2^a > \frac{c 2^{2^{k+1}+2k} m}{(r/\log|F|)^{1/6}}$. Then for any $C$ and $M = c'(r/\log_2 |F|)^{1/6}/(C2^k)$,*

- *any $T^{cc}(k+1, C)$ refutation of $G$ of rank $R$ must have $R \log_2 R \geq M$, and*
- *any tree-like $T^{cc}(k+1, C)$ refutation of $G$ of size $S$ must have $\log S \log \log S \geq M$.*

On the other hand, one can upper bound the rank complexity of $\text{Lift}_{k,a}^{\oplus}(F)$ in terms of that of $F$, even in resolution.

**Lemma 3.11.** *Let $F$ be a $t$-CNF formula on $m$ variables having resolution rank $r$. There is some absolute constant $c > 0$ such that for any $a \geq 1$, there is a resolution refutation of $G = \text{Lift}_{k,a}^{\oplus}(F)$ of rank at most $crka$.*

*Proof.* It is straightforward to construct a decision tree for $G_{\text{search}}$ given one for $F_{\text{search}}$. Whenever a variable $e_i$ in $F$ is queried, the decision tree for $G_{\text{search}}$ makes $ka$ queries to the $ka$ variables in $Y_i$ to find the selected $x_{i,c}$ whose value replaces $e_i$ in evaluating $G$. Thus the depth is multiplied by $O(ka)$. □

**Corollary 3.12.** *Let $t$ be some constant. Suppose that a family of polynomial size $t$-CNF formulas $F$ on $m$ variables has resolution rank complexity $r = r(m)$ that is $m^{\Omega(1)}$. Then, for every $\epsilon > 0$ and $k \geq 1$, there is a family of CNF formulas $G = \text{Lift}_k^{\oplus}(F)$ on $n$ variables of size $n^{O(tk)}$ such that if $k \leq (1-\epsilon) \log \log n$ then*

- *$G$ has $R^{cc}(k+1)$ refutation rank complexity $\Omega(r^{1/7}) = n^{\Omega(1)}$;*
- *there is a resolution refutation of $G$ of rank $O(rk \log n)$;*

- $G$ requires $T^{cc}(k+1)$ tree-size $\exp(n^{\Omega(1)})$.

*Proof.* We apply Theorem 3.10 with $a$ being the least integer satisfying the constraint; that is, $2^a$ is $\frac{c''2^{k+1+2k}m}{(r/\log m)^{1/6}}$ for some constant $c'' > 0$ since $|F|$ is polynomial in $m$.

By Proposition 3.9, $n = m(2^a + ka)$. Now for $k \leq (1-\epsilon)\log\log n$ and since $r$ is $m^{\Omega(1)}$, for sufficiently large $n$ we have $2^a + ka < c''2^{2^{k+2}}m \leq c''n^{1-\delta}m$ for some constant $\delta > 0$. It follows that $n = m(2^a + ka) \leq c''n^{1-\delta}m^2$ and hence $m$ is $n^{\Omega(1)}$. Also by Proposition 3.9, $|G|$ is $|F|2^{tka+ta}(ka+t)$, which is $n^{O(tk)}$.

Suppose that there is a $T^{cc}(k+1)$ refutation $P$ of $G$ of rank $R$. Hence by definition, there is some constant $\beta > 0$ such that $P$ is a $T^{cc}(k+1, \log^\beta n, 1/3)$ refutation. By Theorem 3.10, we have $R \log R = \Omega((r/\log m)^{1/6}/(2^k \log^\beta n))$. Thus for sufficiently large $n$, $R = \Omega(r^{1/7}) = n^{\Omega(1)}$ since $r$ is $m^{\Omega(1)}$.

The rank upper bound follows easily from Lemma 3.11 and the proof for the tree-like size lower bound is similar. □

In particular, by Proposition 2.5, Corollary 3.12 applies to all $Th(k)$ proof systems.

## 4 Rank and Tree-like Size Separations of the Proof System Hierarchy

In this section we separate $R^{cc}(k)$ and $CP(k)$ in terms of rank and tree-like size, thereby separating $R^{cc}(k+1)$ from $R^{cc}(k)$ and $T^{cc}(k+1)$ from $T^{cc}(k)$. The main idea is that if an unsatisfiable $t$-CNF formula $F$ has a small rank CP proof, then we will show that $G = \text{Lift}_{k-1,\ell}^T(F)$ has a small rank $CP(k)$ proof (that can be made small and tree-like). Moreover, if $F$ requires large resolution rank, then with the right parameters, $G$ has no small rank (or small tree-like) $T^{cc}(k)$ proof. Thus $G$ is a separating instance.

The pigeonhole principle is known to be hard for resolution but admits a small rank CP proof. Since we need the clauses of the input formula to be of constant size for the size of the formula $\text{Lift}_{k,\ell}^T(F)$ to be polynomial, we use the following generalization of the pigeonhole principle [5].

Let $\mathcal{G} = (U \cup V, E)$ be any bipartite graph, where $U$ represents the pigeons and $V$ the holes and associate a variable $0 \leq e_{(u,v)} \leq 1$ with each edge $(u,v) \in E$. $\mathcal{G}$−PHP consists of the following clauses, which have been translated to inequalities:

(P) for all $u \in U$: $\sum_{(u,v)\in E} e_{(u,v)} \geq 1$

(H) for all $u \neq u' \in U, v \in V$ s.t. $(u,v),(u',v) \in E$: $e_{(u,v)} + e_{(u',v)} \leq 1$

**Proposition 4.1** ([5]). *For every $n$, there is a bipartite graph $\mathcal{G} = (U \cup V, E)$, where $|U| = |V|+1 = n$ and the degree of every vertex in $U$ is $\leq 5$, such that $\mathcal{G}$−PHP is a polynomial size 5-CNF on $m = 5n$ variables and requires resolution rank $\Omega(m)$.*

From this we immediately obtain a rank lower bound for a lifting of $\mathcal{G}$−PHP.

**Lemma 4.2.** *There is a family of bipartite graphs $\mathcal{G}$ and a family of polynomial-size CNF formulas $\text{Lift}_{k-1}^T(\mathcal{G}\text{−PHP})$ on $n$ variables that requires refutation rank $n^{\Omega(1/k)}$ and tree-like refutation size $\exp(n^{\Omega(1/k)})$ in any $R^{cc}(k)$ systems for any $k \leq (1-\epsilon)\log\log n$ where $\epsilon > 0$ is some absolute constant.*

*Proof.* Let $\mathcal{G}$ be as given by Proposition 4.1. Then $\mathcal{G}-$PHP has linear resolution rank. The lemma follows from Corollary 3.8. □

Our upper bound for the lifted versions of $\mathcal{G}-$PHP will be derived from the following CP rank upper bound for $\mathcal{G}-$PHP itself.

**Proposition 4.3** ([7]). *For any $\mathcal{G} = (U \cup V, E)$ with $|U| = |V| + 1$, $\mathcal{G}-$PHP has a CP refutation of rank $O(\log |U|)$.*

Before considering the lifted versions of $\mathcal{G}-$PHP directly, we first give a generic method for easily deriving some convenient $\text{CP}(k)$ consequences for lifted formulas. Suppose that $F$ has variables $e_1, \ldots, e_m$ and let $G = \text{Lift}_{k-1}^{\text{T}}(F)$. The variables in $G$ are $x_{i,c}$ (recall that each cell $c$ is indexed by a tuple in $[\ell]^{k-1}$) and $y_{i,p,a}$, where $1 \leq i \leq m$, $1 \leq p \leq k-1$, and $1 \leq a \leq \ell$.

For each variable $e_i$ of $F$ define a degree $k$ polynomial $\mathbf{e_i}$ as

$$\mathbf{e_i} := \sum_{c=(a_1,\ldots,a_k)\in[\ell]^{k-1}} x_{i,c}\mathbf{y_{i,c}},$$

where

$$\mathbf{y_{i,c}} := y_{i,1,a_1} \cdot y_{i,2,a_2} \cdots y_{i,k-1,a_{k-1}}.$$

We show how to convert the original axiom clauses (I), (II), and (III) in $G$ into the following forms that are easier to manipulate in $\text{CP}(k)$ systems:

(I') for all $1 \leq i \leq m$: $\sum_{c \in [\ell]^{k-1}} \mathbf{y_{i,c}} \geq 1$

(II') for all $1 \leq i \leq m$ and $c \neq c' \in [\ell]^{k-1}$: $\mathbf{y_{i,c}} + \mathbf{y_{i,c'}} \leq 1$

(III') for all clauses in $F$, say $\neg e_{i_1} \vee e_{i_2} \vee \cdots \vee e_{i_t}$, and for every $t$-tuple of cells $(c^1, \ldots, c^t)$,

$$\mathbf{y_{i_1,c^1}}x_{i_1,c^1} + \mathbf{y_{i_2,c^2}}(1 - x_{i_2,c^2}) + \cdots + \mathbf{y_{i_t,c^t}}(1 - x_{i_t,c^t}) \leq t - 1$$

**Lemma 4.4.** *For any CNF formulas $F$ and $G = \text{Lift}_{k-1,\ell}^{\text{T}}(F)$ for any $k, \ell \geq 2$, there are $\text{CP}(k)$ derivations of rank $k$ of all (I'), (II'), and (III') inequalities as well as $0 \leq \mathbf{y_{i,c}} \leq 1$ and $0 \leq \mathbf{e_i} \leq 1$ given the families of clauses (I), (II), and (III) in $G$.*

*Proof.* Note that the $\text{CP}(k)$ rule that $(q \geq 0) \vdash (x_i q \geq 0)$ for all polynomials $q$ of degree at most $k-1$ and variables $x_i$ implies that if we have inequalities $q_1 \geq b_1$ and $q_2 \geq b_2$ such that the sum of the degrees of $q_1$ and $q_2$ is at most $k$ then $q_1 q_2 \geq b_1 b_2$ is derivable in $\text{CP}(k)$ in rank at most the minimum of the degrees of $q_1$ and $q_2$.

The facts that $0 \leq \mathbf{e_i}$ and $0 \leq \mathbf{y_{i,c}}$ then follow immediately in rank $k-1$ from $0 \leq y_{i,p,a}$ and $0 \leq e_i$.

The (I) axioms in $G$ of the form $\vee_{a \in [\ell]} y_{i,p,a}$ translate to $\sum_{a \in [\ell]} y_{i,p,a} \geq 1$. Applying this product rule of $\text{CP}(k)$ for $p \in [k-1]$ we multiply all of the inequalities together in total rank $k-2$ to obtain inequality (I') above. By the product rule and $x_{i,c} \leq 1$, in rank $k-1$ we obtain $x_{i,c}\mathbf{y_{i,c}} \leq \mathbf{y_{i,c}}$ for all $i, c$ and combining with (I') we obtain that $\mathbf{e_i} \leq 1$ for all $i$.

To obtain an inequality of type (II'), consider some index $j$ such that $c_p \neq c'_p$. We have the translation of the (II) axiom of $G$ ($\neg y_{i,p,c_p} \vee \neg y_{i,p,c'_p}$) which yields $y_{i,p,c_p} + y_{i,p,c'_p} \leq 1$. Since we also have $y \leq 1$ for every variable $y$, by applying the product rule $k-2$ times we have $\mathbf{y_{i,c}} \leq y_{i,p,c_p}$ and $\mathbf{y_{i,c'}} \leq y_{i,p,c'_p}$ and thus (II') follows immediately. (The weaker constraint that $\mathbf{y_{i,c}} \leq 1$ is also an immediate implication.)

To obtain an inequality of type (III'), we use the translation of the (III) clauses of $G$ of the form $\vee_{j=1}^{t}(\vee_{p=1}^{k-1} \neg y_{i_j,p,c_p^j} \vee x_{i_j,c^j})$ which is $\sum_{j=1}^{t}(\sum_{p=1}^{k-1}(1 - y_{i_j,p,c_p^j}) + x_{i_j,c^j}) \geq 1$. Observe that in rank $k$ by the product rule and $0 \leq x_{i_j,c^j} \leq 1$ we can derive

$$\sum_{p=1}^{k-1} y_{i_j,p,c_p^j} + (1 - x_{i_j,c^j}) \geq k(1 - x_{i_j,c^j}) \prod_{p=1}^{k-1} y_{i_j,p,c_p^j} = k(1 - x_{i_j,c^j})\mathbf{y_{i_j,c^j}}.$$

Therefore we have

$$\sum_{p=1}^{k-1}(1 - y_{i_j,p,c_p^j}) + x_{i_j,c^j} \leq k - k(1 - x_{i_j,c^j})\mathbf{y_{i_j,c^j}}.$$

Plugging this into the original inequality we obtain that $\sum_{j=1}^{t}(k - k(1 - x_{i_j,c^j})\mathbf{y_{i_j,c^j}}) \geq 1$. Dividing everything by $k$ and rounding up yields $\sum_{j=1}^{t}(1 - (1 - x_{i_j,c^j})\mathbf{y_{i_j,c^j}}) \geq 1$. Rewriting, we derive $\sum_{j=1}^{t}(1 - x_{i_j,c^j})\mathbf{y_{i_j,c^j}} \leq t - 1$ which is the corresponding inequality (III').

Thus in rank $k$ we can derive all the inequalities (I'), (II'), and (III'). $\square$

We now have the tools to derive an upper bound on the CP($k$) rank of lifted $\mathcal{G}$−PHP formulas and complete the rank separation.

**Lemma 4.5.** *For any $\mathcal{G} = (U \cup V, E)$ with $|U| = |V| + 1$ and the degree of every vertex in $U$ is at most $t$, $G = \text{Lift}_{k-1,\ell}^{\text{T}}(\mathcal{G}-\text{PHP})$ has a CP($k$) refutation of rank $O(\log |U| + tk \log \ell)$, for any $k, \ell \geq 2$.*

*Proof.* For ease of notation, we denote the variables in $F = \mathcal{G}-\text{PHP}$ as $e_1, \ldots, e_m$ where $m = |E|$. The idea is that we will simulate the CP-refutation for $F$ in Proposition 4.3 by replacing each variable $e_i$ in the proof with the degree $k$ polynomial $\mathbf{e_i}$ (together with the associated degree $k - 1$ polynomials $\mathbf{y_{i,c}}$) using the inequalities (I'), (II'), (III') and $0 \leq \mathbf{e_i} \leq 1$ from Lemma 4.4. The rank of the new refutation given these degree $k$ inequalities will be the same as that of the CP-refutation of $F$.

By Lemma 4.4 there are rank-$k$ derivations of all the axiom inequalities in $F$ (consisting of $0 \leq e_i \leq 1$ and (P) and (H) inequalities) with $e_i$ replaced with $\mathbf{e_i}$, given the original axiom clauses (I), (II), and (III) in $G$ (as defined in Section 3.2) so there will be a CP($k$)-refutation of $G$ of rank $k + r$, where $r$ is the rank of the CP-refutation of $F$.

**Claim 4.6.** *Given inequalities (I'),(II'), and (III'), for all (P)-type axioms $e_{i_1} + \ldots + e_{i_t} \geq 1$, for some $t > 0$, in $F = \mathcal{G}-\text{PHP}$, the inequality $\mathbf{e_{i_1}} + \ldots + \mathbf{e_{i_t}} \geq 1$ is rank-$O(tk \log \ell)$ derivable in CP($k$).*

*Proof.* Denoting $\mathbf{z_{i,c}} := x_{i,c}\mathbf{y_{i,c}}$, our goal is to derive

$$\sum_{c \in [\ell]^{k-1}} \mathbf{z_{i_1,c}} + \cdots + \sum_{c \in [\ell]^{k-1}} \mathbf{z_{i_t,c}} \geq 1,$$

given that, from type (III') inequalities, for every $t$-tuple of cells $(c^1, \ldots, c^t)$,

$$\mathbf{y_{i_1,c^1}} + \ldots + \mathbf{y_{i_t,c^t}} \leq t - 1 + \mathbf{z_{i_1,c^1}} + \ldots + \mathbf{z_{i_t,c^t}},$$

and, from type (I') axioms, for each $i \in \{i_1, \ldots, i_t\}$,

$$\sum_{c \in [\ell]^{k-1}} \mathbf{y_{i,c}} \geq 1.$$

We will proceed in $t$ steps, where at step $j$, we will derive, for every $t-j$-tuple of cells $(c^{j+1}, \ldots, c^t)$,

$$(S_j:) \ \mathbf{y_{i_{j+1},c^{j+1}}} + \ldots + \mathbf{y_{i_t,c^t}} \leq t - j - 1 + \left(\sum_{c \in [\ell]^{k-1}} \mathbf{z_{i_1,c}} + \cdots + \sum_{c \in [\ell]^{k-1}} \mathbf{z_{i_j,c}}\right) + (\mathbf{z_{i_{j+1},c^{j+1}}} + \ldots + \mathbf{z_{i_t,c^t}}).$$

Thus we will be done at the end of step $t$. Now, to proceed by induction, assuming we have finished step $j$ and now at step $j+1$. First, for every $t-j-1$-tuple of cells $(c^{j+2}, \ldots, c^t)$, we add together all $S_j$-inequalities for all cells $c^{j+1} \in [\ell]^{k-1}$ and replace $\sum_{c \in [\ell]^{k-1}} \mathbf{y_{i_{j+1},c}} \geq 1$ to get

$$1 + \ell^{k-1}(\mathbf{y_{i_{j+2},c^{j+2}}} + \ldots + \mathbf{y_{i_t,c^t}}) \leq \ell^{k-1}(t-j-1) + \sum_{c \in [\ell]^{k-1}} \mathbf{z_{i_{j+1},c}}$$
$$+ \ell^{k-1}\left(\sum_{c \in [\ell]^{k-1}} \mathbf{z_{i_1,c}} + \cdots + \sum_{c \in [\ell]^{k-1}} \mathbf{z_{i_j,c}}\right) + \ell^{k-1}(\mathbf{z_{i_{j+2},c^{j+2}}} + \ldots + \mathbf{z_{i_t,c^t}}).$$

Next we add $\ell^{k-1} - 1$ copies of $\sum_{c \in [\ell]^{k-1}} \mathbf{z_{i_{j+1},c}}$ to the right side and divide by $\ell^{k-1}$ to get an $S_{j+1}$ inequalities. Each step requires rank $O(k \log \ell)$ with fan-in 2. The claim follows. □

**Claim 4.7.** *For all (H) axioms $e_{i_1} + e_{i_2} \leq 1$ in $F$, $\mathbf{e_{i_1}} + \mathbf{e_{i_2}} \leq 1$ is rank-$O(k \log \ell)$ derivable in* CP($k$).

*Proof.* For $i$ be either $i_1$ or $i_2$ and for any $c \neq c' \in [\ell]^{k-1}$, in one step we can derive $\mathbf{y_{i,c}} x_{i,c} + \mathbf{y_{i,c'}} x_{i,c'} \leq 1$ from the (II') inequality $\mathbf{y_{i,c}} + \mathbf{y_{i,c'}} \leq 1$.

For every pair of cells $(c_1, c_2)$, we are also given the type (III') inequality $\mathbf{y_{i_1,c_1}} x_{i_1,c_1} + \mathbf{y_{i_2,c_2}} x_{i_2,c_2} \leq 1$.

We need to derive $\sum_{c \in [\ell]^{k-1}} x_{i_1,c} \mathbf{y_{i_1,c}} + \sum_{c \in [\ell]^{k-1}} x_{i_2,c} \mathbf{y_{i_2,c}} \leq 1$. Thus we want that the sum of a set of $O(\ell^{k-1})$ variables to be at most 1, given that the sum of any pair of them is at most 1. By a result of ([7], Theorem 6.1), this can be done in rank $O(k \log \ell)$. □

Lemma 4.5 follows from Claims 4.6 and 4.7 □

Putting Lemmas 4.2 and 4.5 together we obtain the following separations of our proof system hierarchy.

**Theorem 4.8.** *For any $\epsilon > 0$ there is a family of unsatisfiable CNF formulas $G$ on $n$ variables that requires nearly polynomial refutation rank $n^{\Omega(1/\log \log n)}$ and nearly exponential tree-like size $\exp(n^{\Omega(1/\log \log n)})$ in all $T^{cc}(k)$ systems but has logarithmic refutation rank and polynomial tree-like refutation size in* CP($k$) *systems for any $k \leq (1-\epsilon) \log \log n$.*

*Proof.* The bounds for rank and the lower bounds for tree-like size follow immediately from Lemmas 4.2 and 4.5. The upper bound for tree-like size follows by expanding the logarithmic rank CP($k$) proof into a tree. □

## 5 Integrality Gaps

In this section we how how to use this approach to obtain not just rank lower bounds for unsatisfiable formulas, but integrality gaps for optimization problems as well. We will present an integrality gap for MAX-SAT as a canonical example.

The MAX-SAT problem is well-studied in the theory of approximation algorithms and optimal inapproximability results are known under the assumption that $\mathsf{P} \neq \mathsf{NP}$. There are also unconditional inapproximability results known for a restricted class of algorithms that involve applying Cutting Planes or LS+ procedures to a relaxation of the standard integer program (e.g. [7, 40]).

Given a CNF formula $G = \{C_1 \wedge \cdots \wedge C_m\}$ over variables $x_1, \ldots, x_n$, we can add a new set of variables $z_1, \ldots, z_m$, and define $C_i' = \neg z_i \vee C_i$. Let $G'$ be $C_1' \wedge \cdots \wedge C_m'$. If we convert these clauses into linear constraints and add Boolean constraints, we obtain a linear program $L_G$ with objective function $\sum_i z_i$ that is a natural LP relation of the MAX-SAT problem for $G$.

There are $2^t \binom{n}{t}$ clauses over $n$ variables that contain exactly $t$ different variables. Let $\mathcal{N}_m^{t,n}$ be the probability distribution induced by choosing $m$ of these clauses uniformly and independently.

Let $F$ be a $t$-CNF formula. We consider $G = \text{Lift}_{1,2}^{\mathrm{T}}(F)$ as described earlier, except that since we have set $k = 1$ and $\ell = 2$, the form of $G$ can be considerably simplified. That is, the variables of $G$ will consist of two bit-vectors, $x$ and $y$, in which $x$ will contain $n$ blocks, each of size 2, $y$ will be a vector of length $n$, where $y_i$ indicates which of the two elements of block $i$ will be chosen in $x$. Each clause of $F$ is transformed into $2^t$ clauses in $G$, corresponding to each of the $2^t$ possible bits of $x$ that could be chosen by $y$. Thus if $F$ has $m$ clauses, each of size $t$, $G$ has $2^t m$ clauses, each of size $2t$, and $F$ is unsatisfiable if and only if $G$ is unsatisfiable. (There is no need for the clauses on the $y_i$-variables that were used in the case of larger $\ell$.)

The following theorem, which is key to getting an integrality gap, is a quantitatively stronger version of Theorem 3.2, for the case $k = 1$ and $\ell = 2$.

**Theorem 5.1.** *Let $F$ be any $t$-CNF formula on $n$ variables having resolution rank at least $r$, and let $G = \text{Lift}_{1,2}^{\mathrm{T}}(F)$. Then any $R^{cc}(2, C)$ refutation of $G$ of rank $R$ must have $CR \log R \geq r^\delta$ for some constant $\delta > 0$.*

In order to prove Theorem 5.1, we will rely on the following stronger version of Theorem 3.2 for the special case of 2 players due to Sherstov [46]. The version we have already seen requires that $\ell$ be large, which becomes a problem for obtaining integrality gaps. The theorem below has much less dependence on the degree, and as a result it does not require $\ell$ to be large. However, this quantitatively stronger version below is currently only known to hold for 2-player communication complexity.

**Theorem 5.2.** *[46] Let $f$ be a boolean function on $n$ variables with sign-degree at least $d$ (and hence $5/6$-degree at least $d$). Then any 2-party communication protocol $\mathcal{P}$ computing $g = f \circ \psi_{1,2}^{\mathrm{T}}$ with error $1/3$ must have $|\mathcal{P}| = \Omega(d)$.*

*Proof of Theorem 5.1.* We proceed as in the proof of Theorem 3.6. Suppose that $P$ is a $R^{cc}(2, C)$ refutation of $G = \text{Lift}_{1,2}^{\mathrm{T}}(F)$ of rank $R$. By Lemma 2.8, there exists a 2-party protocol $\mathcal{P}$ consistently computing $G_{\text{search}}$ of error $1/3$ such that $|\mathcal{P}|$ is $O(CR \log R)$. Now on the one hand, by Lemma 3.5, there exists a consistent system $\mathcal{Z} = \{f_S : S \subseteq \text{clauses}(F)\}$ of functions for $F$ such that for every $S$, there exists a 2-party protocol $\mathcal{P}_S$ computing $f_S \circ \psi_{1,2}^{\mathrm{T}}$ of error $1/3$ such that $|\mathcal{P}_S|$ is $O(CR \log R)$.

On the other hand, by Propositions 2.2 and 2.3, we have

$$D(f_S) \geq \frac{D(F_{\text{search}})}{\lceil \log_2 |F| \rceil} = \frac{r}{\lceil \log_2 |F| \rceil}.$$

Finally by Proposition 3.1, we have $d = deg_{5/6}(f_S) \geq (D(f_S))^{1/6}/4 \geq (\frac{r}{\log_2 |F|})^{1/6}/4$.

Now by the above Theorem 5.2, we must have $CR \log R$ that is $\Omega(d)$, which is $r^\delta$ for some constant $\delta < 1$. □

We now see how the above theorem can be applied to derive an integrality gap for small rank Th(1) or Cutting Planes proofs.

**Corollary 5.3.** *Let $t \geq 3$ be an integer. There exists $\delta < 1$ such that for all $\epsilon > 0$ there is a $\Delta > 1$ such that for a randomly chosen $F$ from $\mathcal{N}_{\Delta n}^{t,n}$, the integrality gap of any $n^\delta$ round Cutting Planes (or Th(1)) relaxation of $L_G$, the linear relaxation of the $2t$-CNF $G = \text{Lift}_{1,2}^T(F)$, is at least $1 - 1/2^{2t} + \epsilon$ with high probability.*

*Proof.* Given $\epsilon$, fix $\Delta >> 2^t \ln 2/\epsilon'^2$, where $(1-1/2^t+\epsilon')(2^t/(2^t-1)-\epsilon) = 1$. A random assignment satisfies each of $F$'s clauses with probability $1 - 1/2^t$, so the expected number of satisfied clauses of $F$ is $(1-1/2^t)\Delta n$. For appropriate choice of $\Delta$, the probability that a random assignment satisfies more than a $1 - 1/2^t + \epsilon'$ fraction of equations is less than $2^{-n}$ by Chernoff bounds. Thus with high probability, no assignment satisfies more than a $1 - 1/2^t + \epsilon'$ fraction of $F$'s equations. By the construction of $G$ from $F$, each clause of $F$ has precisely $2^t$ corresponding clauses in $G$. It follows that with high probability, no assignment satisfies more than a $1 - 1/2^{2t} + \epsilon$ fraction of $G$'s clauses.

On the other hand, since $t \geq 3$, any Resolution refutation of $F$ requires linear rank [13, 5]. Thus by Theorem 5.1 even after Th(1) inference of rank $n^\delta$ (and in particular $n^\delta$ rounds of Cutting Planes), there is some non-integral assignment $\alpha$ to the $x_i's$ that satisfies all linear constraints corresponding to the clauses of $G$. Extend this assignment by setting all the $z_i$'s to 1 and it follows that all constraints of $L_G$, are also satisfied. Thus we have a solution satisfying all equations that survives even after $n^\delta$ rounds. □

Note that since a random assignment on average satisfies a $1 - 1/2^{2t}$ fraction of clauses of any $2t$-CNF formula, this yields an optimal integrality gap for rank $n^\delta$ Th(1) inference for MAX-$2t$-SAT for any $t \geq 3$. Such a result was previously known for the special case of Cutting Planes proofs [7] but the proof relied on the specific form of inference rather than the general sound inference allowed for Th(1) proofs.

**Remark.** We can obtain a similar integrality gap for *any* function where we can prove decision tree lower bounds. That is, take any optimization problem that can be expressed naturally as a $t$-CNF formula, and such that an integrality gap of $1 - \gamma$ can be proven for decision trees. (Such a result is usually elementary to obtain.) Then by our lifting technique, we can show that any small $R^{cc}(2,C)$ refutation (including Cutting Planes and Th(1) proofs) for the lifted version has an integrality gap of $1 - \gamma/2^t$. Our approach only works at present for proof systems that correspond to 2-player communication complexity. However, an extension of Theorem 5.2 to the multiparty setting (as was done with the qualitatively weaker Theorem 3.2 which was originally proven for the 2-player case [45]) would immediately yield integrality gaps for stronger matrix cut systems, such as LS+ and Lasserre.

## 6  Discussion

In this paper we showed how to take an arbitrary 3-CNF formula $F$, and convert it into another CNF formula $G$ so that the resolution rank of $F$ becomes polynomial in the $T^{cc}(k+1)$ rank of $G$, for $k \in O(\log \log n)$. As applications, we obtained polynomial rank lower bounds for many commonly studied matrix cut proof systems, including Cutting Planes and the full complement of Lovász-Schrijver variants, as well as non-constant rank lower bounds for Sherali-Adams and Lasserre proofs. We also use our approach to obtain new hierarchy theorems for the systems $CP(k)$ and $LS^k_{+,*}$.

While we focused on semi-algebraic systems in this paper, we would like to point out that our theorems can also be used to obtain non-constant rank lower bounds for many commonly studied algebraic systems, including Hilbert's Nullstellensatz and the Polynomial Calculus. While stronger lower bounds for these latter systems were already known prior to our work, our method achieves these lower bounds for a large class of new CNF formulas and, the technique is simple and generic. It should also be possible to obtain degree-based hierarchy theorems using our approach.

There are several interesting open problems directly related to our work. First, our theorems as stated work for $k$ up to $(1 - o(1)) \log \log n$. We conjecture that it should be possible to derive hardness escalation results that work for $k$ up to $\Omega(\log n)$. A key problem with our approach is the tensor selector method, which when applied for larger $k$, introduces superpolynomially many variables. A similar problem arose when proving lower bounds set disjointness and related functions in the NOF communication model. The initial results ([31, 11]) used the tensor selector and worked for $k = (1 - o(1)) \log \log n$; subsequent papers introduced new selector methods in order to prove lower bounds for $k = \Omega(\log n)$ ([14, 3].) On the other hand, proving hardness escalation results for $k = \omega(\log n)$ appears to require very new ideas and would solve a major open problem in communication complexity and circuit complexity.

Secondly, an important open problem is to strengthen our method to obtain not only tree-size lower bounds, but general (dag-like) size lower bounds. We note that this has already happened for $k = 2$, where initially Cutting Planes tree-size lower bounds were proven based on two-player communication complexity lower bounds [24] and the results were later generalized to obtain unrestricted Cutting Planes size lower bounds [6, 37]. Such a result, even for $k = 3$, would give unrestricted size lower bounds for Lovász-Schrijver proofs, thus solving an important open problem.

Finally, there are many very interesting questions related to hardness escalation. What relationships are there between these various forms of hardness amplification, hardness escalation, hardness condensing, and hardness amplification? Are there other examples of hardness escalation, even under reasonable assumptions? In particular it would be very interesting to obtain a hardness escalation result that lifts lower bounds for a circuit class where cryptography is not possible to a circuit class were cryptography is possible (e.g., lifting from DNF lower bounds to $TC_0$ lower bounds) as such a result would cross the "natural proof" barrier.

## Acknowledgements

We would like to thank Ran Raz and Rahul Santhanam for very helpful conversations. In particular we thank Rahul for suggesting the term *hardness escalation*.

# References

[1] Robert Beals, Harry Buhrman, Richard Cleve, Michele Mosca, and Ronald de Wolf. Quantum lower bounds by polynomials. *Journal of the ACM*, 48(4):778–797, 2001.

[2] P. Beame, T. Pitassi, and N. Segerlind. Lower bounds for Lovász-Schrijver systems and beyond follow from multiparty communication complexity. *SIAM Journal on Computing*, 37(3):845–869, 2007.

[3] Paul Beame and Dang-Trinh Huynh-Ngoc. Multiparty communication complexity and threshold circuit complexity of $AC^0$. In *Proceedings of the 50th Annual Symposium on Foundations of Computer Science*, Atlanta,GA, October 2009. IEEE. To appear.

[4] E. Ben-Sasson and J. Nordstrom. Understanding space in resolution: Optimal lower bounds and exponential tradeoffs. Technical Report TR09-034, Electronic Colloquium in Computation Complexity, http://www.eccc.uni-trier.de/eccc/, 2009.

[5] E. Ben-Sasson and A. Wigderson. Short proofs are narrow – resolution made simple. *Journal of the ACM*, 48(2):149–169, 2001.

[6] M. L. Bonet, T. Pitassi, and R. Raz. Lower bounds for cutting planes proofs with small coefficients. *Journal of Symbolic Logic*, 62(3):708–728, September 1997.

[7] J. Buresh-Oppenheim, N. Galesi, S. Hoory, A. Magen, and T. Pitassi. Rank bounds and integrality gaps for cutting planes procedures. In *Proceedings 44th Annual Symposium on Foundations of Computer Science*, pages 318–327, Boston, MA, October 2003. IEEE.

[8] J. Buresh-Oppenheim and T. Pitassi. The complexity of resolution refinements. *Journal of Symbolic Logic*, 72(4):1336–1352, 2007.

[9] J. Buresh-Oppenheim and R. Santhanam. Making hard problems harder. Technical Report TR06-03, Electronic Colloquium in Computation Complexity, http://www.eccc.uni-trier.de/eccc/, 2006.

[10] M. Charikar, K. Makarychev, and Y. Makarychev. Integrality gaps for sherali-adams relaxations. In *Proceedings of the Forty-First Annual ACM Symposium on Theory of Computing*, pages 283–292, Bethesda, MD, May 2009. ACM.

[11] A. Chattopadhyay and A. Ada. Multiparty communication complexity of disjointness. Technical Report TR08-002, Electronic Colloquium in Computation Complexity, http://www.eccc.uni-trier.de/eccc/, 2008.

[12] V. Chvátal. Edmonds polytopes and a hierarchy of combinatorial problems. *Discrete Mathematics*, 4:305–337, 1973.

[13] V. Chvátal and E. Szemerédi. Many hard examples for resolution. *Journal of the ACM*, 35(4):759–768, 1988.

[14] Matei David, Toniann Pitassi, and Emanuele Viola. Improved separations between nondeterministic and randomized multiparty communication. In *RANDOM 2008, 12th International*

*Workshop on Randomization and Approximization Techniques in Computer Science*, pages 371–384, 2008.

[15] M. Davis, G. Logemann, and D. Loveland. A machine program for theorem proving. *Communications of the ACM*, 5:394–397, 1962.

[16] J. Edmonds, R. Impagliazzo, S. Rudich, and J. Sgall. Communication complexity towards lower bounds on circuit depth. *Computational Complexity*, 10:210–246, 2001.

[17] K. Georgiou, A. Magen, T. Pitassi, and I. Tourlakis. Integrality gaps of $2 - o(1)$ for vertex cover SDPs in the Lovasz-Schrijver hierarchy. In *Proceedings 48th Annual Symposium on Foundations of Computer Science*, pages 702–712, Berkeley, CA, October 2007. IEEE.

[18] K. Georgiou, A. Magen, and M. Tulsiani. Optimal Sherali-Adams gaps from pairwise independence. Technical Report TR09-061, Electronic Colloquium in Computation Complexity, http://www.eccc.uni-trier.de/eccc/, 2009.

[19] R.E. Gomory. Outline of an algorithm for integer solutions to linear programs. *Bulletin of the American Mathematical Society*, 64:275–278, 1958.

[20] D. Grigoriev. Linear lower bound on degrees of Positivstellensatz calculus proofs for the parity. *Theoretical Computer Science*, 259:613–622, 2001.

[21] D. Grigoriev, E. A. Hirsch, and D. V. Pasechnik. Complexity of semi-algebraic proofs. In *(STACS) 2002: 19th Annual Symposium on Theoretical Aspects of Computer Science*, volume 2285 of *Lecture Notes in Computer Science*, pages 419–430, Antibes, France, February 2002. Springer-Verlag.

[22] D. Grigoriev and N. Vorobjov. The complexity of Null- and Positivstellensatz proofs. *Annals of Pure and Applied Logic*, 113:153–160, 2001.

[23] R. Impagliazzo and R. Paturi. On the complexity of $k$-SAT. *Journal of Computer and System Sciences*, 67:367–375, 2001.

[24] R. Impagliazzo, T. Pitassi, and A. Urquhart. Upper and lower bounds on tree-like cutting planes proofs. In *9th Annual IEEE Symposium on Logic in Computer Science*, pages 220–228, Paris, France, 1994.

[25] M. Karchmer, R. Raz, and A. Wigderson. Super-logarithmic depth lower bounds via direct sum in communication complexity. *Computational Complexity*, 5:191–204, 1995.

[26] S. Khot and R. Saket. SDP integrality gaps with local $\ell_1$-embeddability. In *Proceedings of the 50th Annual Symposium on Foundations of Computer Science*, Atlanta,GA, October 2009. IEEE. To appear.

[27] A. Kojevnikov and A. Itsykson. Lower bounds of static Lovasz-Schrijver calculus proofs for Tseitin tautologies. In *Automata, Languages, and Programming: 33rd International Colloquium*, pages 323–334, July 2006.

[28] J. Krajíček. Discretely ordered modules as a first-order extension of the cutting planes proof system. *Journal of Symbolic Logic*, 63(4):1582–1596, 1998.

[29] E. Kushilevitz and N. Nisan. *Communication Complexity*. Cambridge University Press, Cambridge, England ; New York, 1997.

[30] J.B. Lasserre. An explicit SDP relaxation for nonlinear 0-1 programs. In *Proceedings of the 8th International Conference on Integer Programming and Combinatorial Optimization (IPCO 2001), Lecture Notes in Computer Science v. 2081*, pages 293–303. Springer-Verlag, 2001.

[31] T. Lee and A. Shraibman. Disjointness is hard in the multi-party number-on-the-forehead model. In *Proceedings Twenty-Third Annual IEEE Conference on Computational Complexity*, pages 81–91, College Park, Maryland, June 2008.

[32] L. Lovasz and A. Schrijver. Cones of matrices and set-functions and 0-1 optimization. *SIAM J. Optimization*, 1(2):166–190, 1991.

[33] A. Maciel and T. Pitassi. Lower bounds on constant-depth frege proofs with mod connectives modulo a hardness conjecture. In *21st Annual IEEE Symposium on Logic in Computer Science*, pages 189–200, 2006.

[34] S. Muroga. *Threshold logic and its applications*. John Wiley & Sons, 1971.

[35] P. Nguyen. Separating dag-like and tree-like proof systems. In *Annual IEEE Symposium on Logic in Computer Science*, pages 235–244. IEEE, 2007.

[36] N. Nisan and M. Szegedy. On the degree of boolean functions as real polynomials. *Computational Complexity*, 4:301–314, 1994.

[37] P. Pudlák. Lower bounds for resolution and cutting plane proofs and monotone computations. *Journal of Symbolic Logic*, 62(3):981–998, September 1997.

[38] P. Raghavendra and D. Steurer. Integrality gaps for strong SDP relaxations of unique games. In *Proceedings of the 50th Annual Symposium on Foundations of Computer Science*, Atlanta,GA, October 2009. IEEE. To appear.

[39] R. Raz and P. McKenzie. Separation of the monotone NC hierarchy. *Combinatorica*, 19(3):403–435, 1999.

[40] G. Schoenebeck. Linear level lasserre lower bounds for certain k-csps. In *Proceedings 49th Annual Symposium on Foundations of Computer Science*, pages 593–602, Philadelpha, PA, November 2008. IEEE.

[41] G. Schoenebeck, L. Trevisan, and M. Tulsiani. A linear round lower bound for Lovasz-Schrijver SDP relaxations of Vertex Cover. In *Proceedings Twenty-Second Annual IEEE Conference on Computational Complexity*, pages 205–216, 2007.

[42] G. Schoenebeck, L. Trevisan, and M. Tulsiani. Tight integrality gaps for Lovasz-Schrijver LP relaxations of Vertex Cover and Max Cut. In *Proceedings of the Thirty-Ninth Annual ACM Symposium on Theory of Computing*, pages 302–310, 2007.

[43] N. Segerlind and T. Pitassi. Exponential lower bounds and integrality gaps for tree-like Lovasz-Schrijver procedures. In *Proceedings of the Twentieth Annual ACM-SIAM Symposium on Discrete Algorithms*, pages 355–364, New York, NY, January 2009. Society for Industrial and Applied Mathematics.

[44] A. Sherali and W. Adams. A hierarchy of relaxations between the continuous and convex hull representations for zero-one programming problems. *SIAM Journal on Discrete Mathematics*, 3:411–430, 1990.

[45] A. A. Sherstov. Separating AC$^0$ from depth-2 majority circuits. In *Proceedings of the Thirty-Ninth Annual ACM Symposium on Theory of Computing*, pages 294–301, San Diego, CA, June 2007.

[46] A. A. Sherstov. The pattern matrix method for lower bounds on quantum communication. In *Proceedings of the Fortieth Annual ACM Symposium on Theory of Computing*, pages 85–94, Victoria, BC, May 2008.

[47] A. A. Sherstov. The pattern matrix method (journal version). *CoRR*, abs/0906.4291, 2009.

[48] M. Tulsiani. CSP gaps and reductions in the Lasserre hierarchy. In *Proceedings of the Forty-First Annual ACM Symposium on Theory of Computing*, pages 303–312, Bethesda, MD, May 2009. ACM.

[49] A. Urquhart. Regular and general resolution: An improved separation. In *Proceedings of the Eleventh International Conference on Theory and Applications of Satisfiability Testing (SAT 2008)*, Lecture Notes in Computer Science, pages 277–290. Springer-Verlag, 2008.